\documentclass[aps,twocolumn,showkeys,nofootinbib]{revtex4}
\usepackage{graphicx,amsmath,amssymb,amsfonts,latexsym,color,dcolumn,bm,mathtools,dsfont}
\usepackage{verbatim}
\usepackage{epstopdf}
\usepackage[colorlinks=true,hidelinks]{hyperref}
\usepackage[normalem]{ulem} 

\providecommand{\abs}[1]{\left\lvert#1\right\rvert}

\definecolor{blue}{rgb}{0.223,0.223,0.667}
\definecolor{red}{rgb}{0.7,0,0}

\begin{document}

\title{Nonreciprocal microwave signal processing with a Field-Programmable Josephson Amplifier}

\author{F. Lecocq$^{1*}$, L. Ranzani$^2$, G. A. Peterson$^1$, K. Cicak$^1$, R. W. Simmonds$^1$, J. D. Teufel$^1$ and J. Aumentado$^1$\footnote{Corresponding authors: florent.lecocq@nist.gov and jose.aumentado@nist.gov}}
\affiliation{$^1$National Institute of Standards and Technology, 325 Broadway, Boulder, CO 80305, USA}
\affiliation{$^2$Raytheon BBN Technologies, Cambridge, Massachusetts 02138, USA}
\date{\today}

\begin{abstract}
We report on the design and implementation of a Field Programmable Josephson Amplifier (FPJA) ---a compact and lossless superconducting circuit that can be programmed \textit{in situ} by a set of microwave drives to perform reciprocal and nonreciprocal frequency conversion and amplification. In this work we demonstrate four modes of operation: frequency conversion ($-0.5~\mathrm{dB}$ transmission, $-30~\mathrm{dB}$ reflection), circulation ($-0.5~\mathrm{dB}$ transmission, $-30~\mathrm{dB}$ reflection, $30~\mathrm{dB}$ isolation), phase-preserving amplification (gain $>20~\mathrm{dB}$, $1~\mathrm{photon}$ of added noise) and directional phase-preserving amplification ($-10~\mathrm{dB}$ reflection, $18~\mathrm{dB}$ forward gain, $8~\mathrm{dB}$ reverse isolation, $1~\mathrm{photon}$ of added noise). The system exhibits quantitative agreement with theoretical prediction. Based on a gradiometric Superconducting Quantum Interference Device (SQUID) with Nb/Al--AlO$_x$/Nb Josephson junctions, the FPJA is first-order insensitive to flux noise and can be operated without magnetic shielding at low temperature. Due to its flexible design and compatibility with existing superconducting fabrication techniques, the FPJA offers a straightforward route toward on-chip integration with superconducting quantum circuits such as qubits or microwave optomechanical systems.
\end{abstract}

\keywords{reciprocity, microwave amplifier, circulator, parametric amplification, superconducting circuits, frequency conversion, quantum-limited amplification}

\maketitle

\section{Introduction}

Many superconducting quantum circuits rely on microwave photons to measure or couple quantum systems, such as superconducting qubits or micro-mechanical resonators\cite{Devoret2013,Aspelmeyer2013}. The ability to process microwave fields with minimal degradation is crucial to the observation of truly quantum behavior. For example, quantum-limited amplification maximizes measurement fidelity, a crucial metric in quantum computing \cite{Devoret2013,Barends2014,Sun2014}, quantum feedback \cite{Vijay2012,Riste2013,Wilson2015}, observation of quantum trajectories \cite{Murch2013}, and position measurements \cite{Teufel2008}. Similarly, the efficient routing of microwave photons enables long distance entanglement \cite{Roch2014} and is an important tool in proposals for quantum networks \cite{Kimble2008}.

Recent developments in Josephson junction-based parametric amplifiers have led to an order of magnitude improvement in measurement efficiency compared to commercially available high-electron-mobility transistor (HEMT) amplifiers \cite{Yurke1996,Castellanos2008,Bergeal2010}. However, these amplifiers are \emph{reciprocal} devices, \textit{i.e.}, their scattering parameter amplitudes are symmetric under the exchange of source and detector \cite{Deak2012,Ranzani2015,Metelmann2015}. As a consequence, in order to protect the device-under-test (DUT) from amplifier backaction and to control signal flow, they require the use of microwave circulators to separate input signals from amplified output onto different physical ports. These components drastically reduce quantum measurement efficiencies \cite{Roch2012,Kindel2016}. In addition to their intrinsic loss, circulators are relatively large and require large dc magnetic fields, preventing direct integration into modern superconducting circuits.

The limitations outlined above have motivated the development of nonreciprocal, non-magnetic, lossless circuits \cite{Kamal2011,Kamal2014,Abdo2013,Abdo2014,Ranzani2015,Metelmann2015,Kerckhoff2015,Estep2014,Sliwa2015,Macklin2015}. Many of these approaches seek to provide highly efficient routing or amplification solutions that can be tightly integrated with superconducting circuits. For example, Josephson traveling-wave parametric amplifiers were specifically designed to achieve high directional gain over several gigahertz of bandwidth, with a large dynamic range \cite{Macklin2015}. In this work we demonstrate a wider range of nonreciprocal behavior, using parametrically coupled multi-mode circuits to build an interferometer in frequency space\cite{Ranzani2015,Metelmann2015}. Due to the directional phase shift inherent to parametric interaction, different interferences occur in the forward and backward directions. Complex networks of coupled modes can be programmed \textit{in situ} by choosing a set of applied parametric microwave drives, leading to a variety of nonreciprocal scattering parameters. The versatility of this approach was recently demonstrated in a Josephson Parametric Converter \cite{Sliwa2015}. In this work we present a alternative circuit design, based on lumped-element niobium components coupled via a single gradiometric SQUID, insensitive to flux noise. Experimental measurements of both the scattering parameters and noise performance show quantitative agreement with theoretical calculations. Here we focus on cases in which two or three modes are coupled and demonstrate four basic functions: frequency conversion, circulation, phase-preserving amplification, and directional phase-preserving amplification. In the following we discuss the device and measurement setup and then describe each mode of operation. Detailed calculations, noise calibrations, and device fabrication details are given in appendices.

\begin{figure}
	\includegraphics[scale=1]{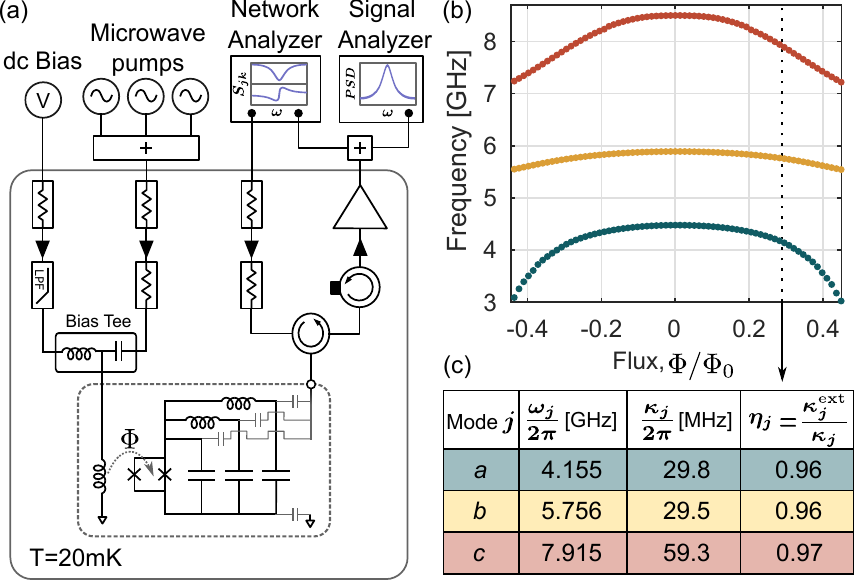}
	\caption{Device and experimental setup. (a) A superconducting resonant circuit is measured in reflection in a cryostat. It exhibits three resonances whose frequencies are tunable by the flux applied to a single SQUID. (b) Measured resonance frequencies (solid lines), $\omega_{a}$, $\omega_{b}$ and $\omega_{c}$, as a function of the flux bias to the SQUID. (c) Table of device parameters for $\Phi/\Phi_0\approx0.29$, showing the resonance frequency $\omega_j$, total linewidth $\kappa_j$, and coupling efficiency $\eta_j=\kappa_j^\text{ext}/\kappa_j$ of each mode $j$.
	\label{figSetup}}
\end{figure}

\section{Description of the FPJA}

The device shown in Fig. \ref{figSetup} consists of three lumped-element resonant circuits in parallel connected to a single SQUID. The SQUID acts as a tunable linear inductor that can be modulated at microwave frequencies. The device is mounted in a dilution refrigerator and measured in reflection, see  Fig. \ref{figSetup}(a). The scattering parameters are measured using a vector network analyzer with a frequency conversion option.  A separate broadband pump line is used to thread flux through the SQUID loop. The circuit exhibits three resonances at frequencies, $\omega_{a}$, $\omega_{b}$ and $\omega_{c}$. Each of these resonances shows a dependence on the SQUID inductance which is in turn modulated by the applied SQUID flux $\Phi$, as shown in Fig. \ref{figSetup}(b). The circuit components were designed to place all three resonances within the $4~\mathrm{GHz}$ to $8~\mathrm{GHz}$ band, while ensuring that all the possible frequency combinations $\omega_j\pm\omega_k$, where $j,k \in \{a,b,c\}$, are well separated ---a critical property for well-controlled parametric interactions (see Appendix B). Coupling capacitors to ground and to the $50~\mathrm{\Omega}$ environment set the external coupling rates to the single measurement line,  $\kappa_{j}^\text{ext}$. The use of a low-loss dielectric for the capacitors (amorphous silicon, loss tangent $\lesssim 5\times10^{-4}$) ensures that these rates exceed the internal loss rates $\kappa_{j}^\text{int}$ by more than an order of magnitude and dominate the total linewidths $\kappa_{j}=\kappa_{j}^\text{ext}+\kappa_{j}^\text{int}$. In the following we will fix the dc flux bias to $\Phi/\Phi_0\approx0.29$, where $\Phi_0$ is a flux quantum, and the measured frequencies and linewidths are summarized in Fig. \ref{figSetup}(c).

\section{Operation of the FPJA}

Currents from the three resonators of the FPJA flow through the SQUID, effectively linearly coupling their dynamics. These dynamics occur at vastly different frequencies, and to first order can be treated independently. However, the modulation of the coupling element, here the SQUID inductance, can lead to parametric coupling between the resonators \cite{Louisell1960}, as discussed in Appendix B. In presence of such a \textit{pump}, the time-dependent coupling strength between modes $j$ and $k$ is 
\begin{equation}
g_{jk}(t)=\frac{\delta\Phi_{jk}(t)}{4}\sqrt{\frac{\partial\omega_{j}}{\partial\Phi}\frac{\partial\omega_{k}}{\partial\Phi}}
\end{equation}  
where $\delta\Phi_{jk}(t)=\abs{\delta\Phi_{jk}}\cos(\omega_{jk}^pt+\phi_{jk})$ is the flux modulation with an amplitude $\abs{\delta\Phi_{jk}}$, a frequency $\omega_{jk}^p$ and phase $\phi_{jk}$. The coupling term in the Hamiltonian of the system depends on the modulation frequency. In particular the pump can mediate two kinds of parametric coupling between two modes. For pump frequencies of the form $\abs{\omega_{j}-\omega_{k}}$, where ${j},{k} \in \{a,b,c\}$, the creation or annihilation of a pump photon enables the coherent exchange of a photon between modes $j$ and $k$, leading to frequency conversion. For pump frequencies of the form $\omega_{j}+\omega_{k}$, the annihilation of a pump photon creates a correlated pair of photons in modes $j$ and $k$, leading to amplification. Importantly, multiple pumps can be simultaneously applied to program an arbitrary set of coupling terms between the modes. A network of modes and couplings can be built, defining the behavior of the circuit. We utilize a graph-based analysis \cite{Ranzani2015} that emphasizes the topology of the coupling network, which is crucial to build intuition and leading to a good agreement with the data. The general methodology consists in solving the Heisenberg-Langevin coupled equations of motions (EoMs) of the circuit, for a given pump configuration. Using input-output formalism we calculate the scattering parameters and the output noise of the system. Detailed calculation are available in Appendix B.

\subsubsection{The FPJA as a frequency converter}

\begin{figure}
	\includegraphics[scale=1]{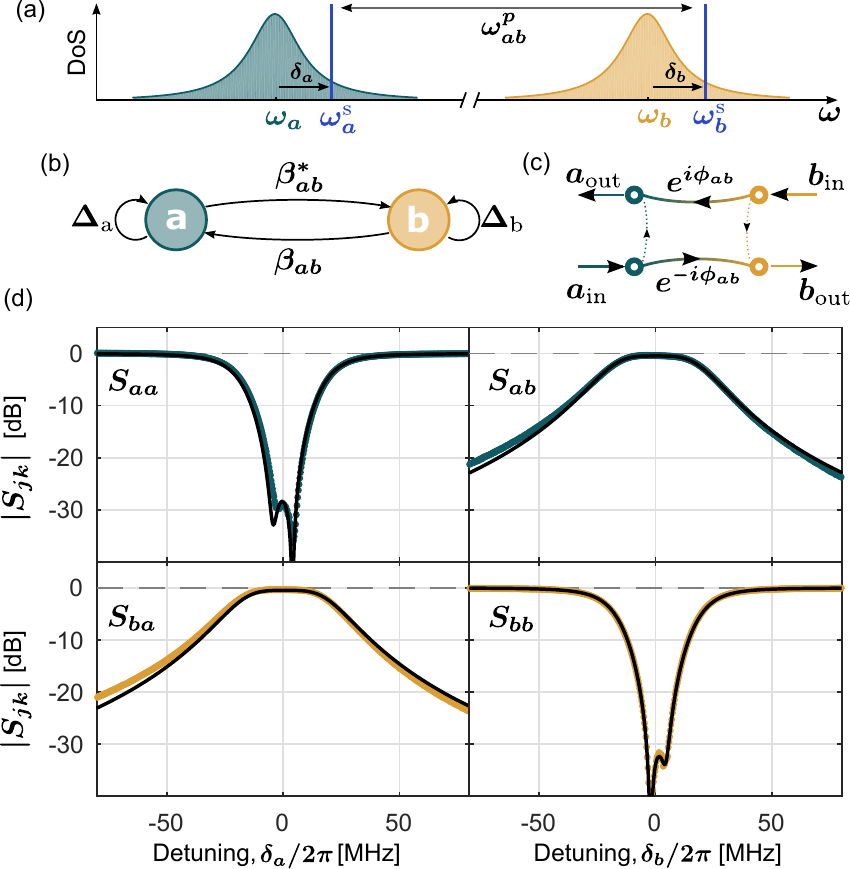}
	\caption{Frequency converter (a) Frequency space diagram: A strong pump of frequency $\omega_{ab}^p = \abs{\omega_{b}-\omega_{a}}$ converts an input signal at the frequency $\omega_{a}^\text{s} = \omega_{a}+\delta_a$ to a output signal frequency $\omega_{b}^\text{s} = \omega_{b}+\delta_b$, and vice versa. (b) Graph representation of the EoM \ref{eq:EoM_FC}. (c) Ideal signal flow diagram. (d) Measured scattering parameters (dots) and fits to Eq. \ref{eq:EoM_FC} (lines), for a fixed pump strength $\abs{\beta_{ab}}\approx0.5$, as a function of the detuning $\delta_j$. The device exhibits good impedance matching (low return loss), and near unity transmission.
		\label{figFC}}
\end{figure}

We start with the first building block: frequency conversion between two modes, here chosen to be $a$ and $b$. We modulate the flux through the SQUID with a pump at a frequency $\omega_{ab}^p\approx \abs{\omega_{b} - \omega_{a}}$. An input signal of amplitude $a_{\text{in}}$, driving the mode amplitude $a$ at a frequency $\omega_{a}^\text{s}$, is coupled to the mode amplitude $b$, leading to an output signal of amplitude $b_\text{out}$ at a frequency $\omega_{b}^\text{s}=\omega_{a}^\text{s}+\omega_{ab}^p$, and vice versa (see Fig. \ref{figFC}(a)). The EoMs in the signal frame reduce to
\begin{equation}
\begin{array}{c}    
\kappa_{a}\Delta_{a}a+\sqrt{\kappa_{a}\kappa_{b}}\beta_{ab}b=i\sqrt{\kappa_{a}^{\text{ext}}}a_{\text{in}},\\
 \\
\kappa_{b}\Delta_{b}b+\sqrt{\kappa_{a}\kappa_{b}}\beta_{ab}^*a=i\sqrt{\kappa_{b}^{\text{ext}}}b_{\text{in}},
\end{array}  
\label{eq:EoM_FC}
\end{equation}
with $\Delta_{j}=(\omega_{j}^\text{s}-\omega_{j})/\kappa_{j}+i/2$ the normalized detuning for mode ${j}$ and $\beta_{ab}=|g_{ab}|e^{i\phi_{ab}}/(2\sqrt{\kappa_a\kappa_b})$ the normalized coupling between the mode amplitudes $a$ and $b$. Importantly, this coupling term is complex, with its phase $\phi_{ab}$ and amplitude $\abs{g_{ab}}$ inherited from the pump. We show in Fig. \ref{figFC}(b) a graph representation of the EoMs in Eq. \ref{eq:EoM_FC}, where vertices represent the mode amplitude and arrows represent the detuning and coupling terms. In the ideal resonant case, defined as $\omega_{ab}^p = \abs{\omega_{b} - \omega_{a}}$ and $\omega_{j}^\text{s}=\omega_{j}$, and neglecting internal loss, $\kappa_{j}^\text{int}=0$, the scattering matrix $\mathbf{S}$ for the system is
\begin{equation}
\mathbf{S}=\left(\begin{array}{cc}
\dfrac{1-4\abs{\beta_{ab}}^2}{1+4\abs{\beta_{ab}}^2} & \dfrac{4i\beta_{ab}}{1+4\abs{\beta_{ab}}^2} \\
\dfrac{4i\beta_{ab}^*}{1+4\abs{\beta_{ab}}^2} & \dfrac{1-4\abs{\beta_{ab}}^2}{1+4\abs{\beta_{ab}}^2}
\end{array}\right)
\label{eq:S_FC_0}
\end{equation}

Close examination of the scattering matrix reveals that unity transmission coincides with impedance matching (zero reflection) at $\abs{\beta_{ab}}=1/2$, and the corresponding signal flow diagram \cite{Pozar2012} is shown in Fig. \ref{figFC}(c). We note here that $S_{ba} = -S_{ab}^*$, na\"{i}vely providing gyration. While this will be crucial later for establishing directionality, on its own this is not unconditionally nonreciprocal due to an ambiguity in the reference frame when describing scattering between two different frequency modes\cite{Leung2010,Deak2012,Ranzani2015}.

The measured scattering parameters in our device are shown in Fig. \ref{figFC}(d) as a function of the detuning $\delta_j=\omega_{j}^\text{s}-\omega_{j}$. We fit the measured response using the full solutions of Eq. \ref{eq:EoM_FC}, finding very good agreement with the data. In particular, we measure only $0.5~\text{dB}$ of insertion loss, fully captured by including the internal loss of the resonators so that $\abs{S_{ba}}^2=\kappa_{a}^\text{ext}\kappa_{b}^\text{ext}/\kappa_{a}\kappa_{b}$ (see Appendix B). The circuit is well matched, with $30~\text{dB}$ of return loss. The bandwidth of conversion is $\sqrt{\kappa_{a}\kappa_{b}}/2\pi\approx30~\text{MHz}$.

\subsubsection{The FPJA as a two mode amplifier}

\begin{figure*}
	\includegraphics[scale=1]{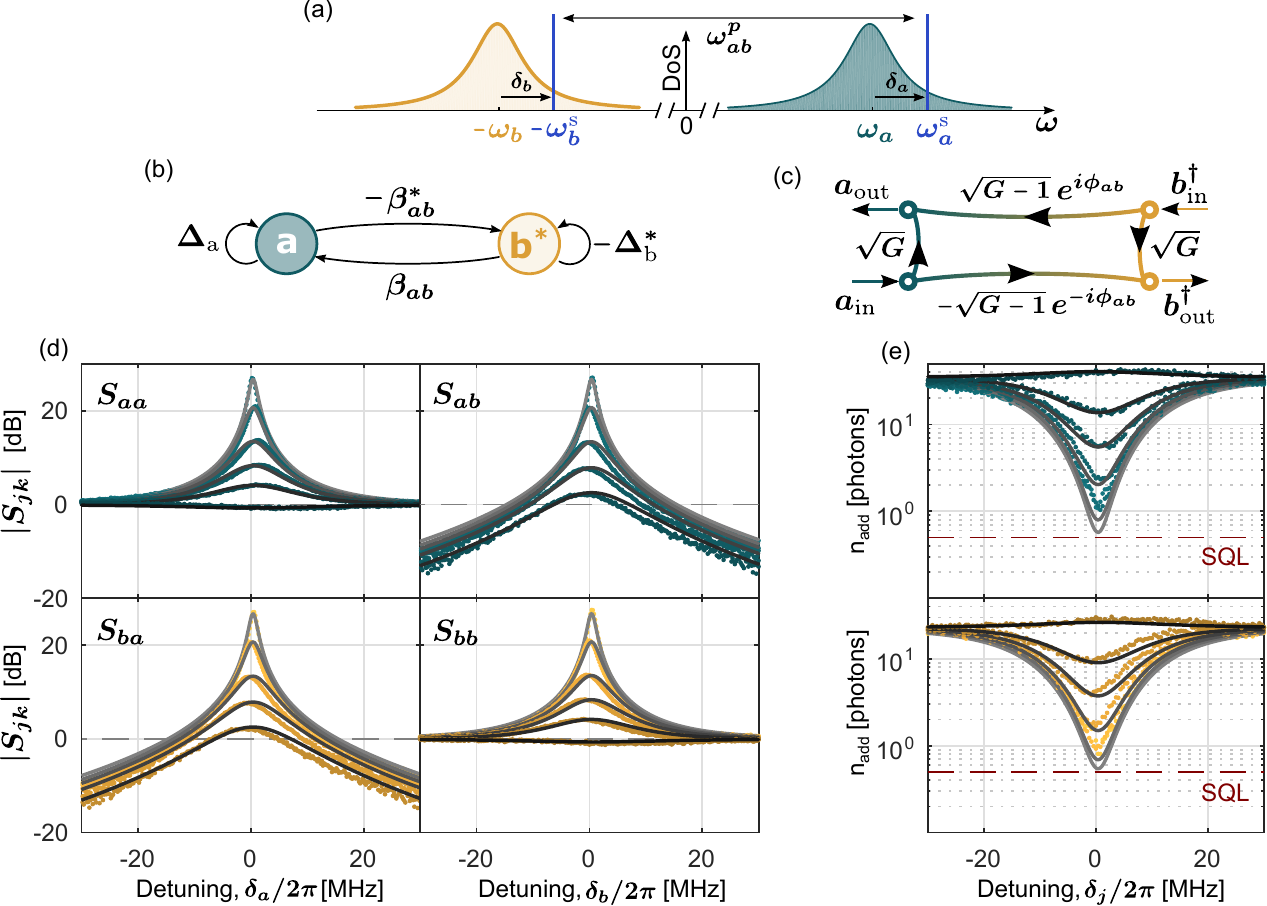}
	\caption{Two-mode amplifier. (a) Frequency space diagram: A strong pump of frequency $\omega_{ab}^p = \omega_{b}+\omega_{a}$ amplifies an input signal at the frequency $\omega_{a}^\text{s} = \omega_{a}+\delta_a$ and generates an idler at the frequency $-\omega_{b}^\text{s} = -\omega_{b}+\delta_b$, and vice-versa. (b) Graph representation of the EoM \ref{eq:EoM_PA} (c) Ideal signal flow diagram (d) Measured scattering parameters (dots) and fits to Eq. \ref{eq:EoM_PA} (lines) as a function of the detuning $\delta_j$, for increasing pump strength $\beta_{ab}$ (dark to light color). (e) Measured system added noise (dots) and theoretical predictions (lines) referred to the input of the FPJA, as a function of the detuning $\delta_j$
	\label{figPA}}
\end{figure*}

We now describe the second building block: amplification between two modes, here chosen to be $a$ and $b$. We modulate the flux through the SQUID with a pump at the sum frequency $\omega_{ab}^p\approx \omega_{b} + \omega_{a}$. An input signal at a frequency $\omega_{a}^\text{s}$ is amplified and generates an idler at $-\omega_{b}^\text{s}=\omega_{a}^\text{s}-\omega_{ab}^p$ (see Fig. \ref{figPA}(a)). In contrast with the frequency conversion case in Eq. \ref{eq:EoM_FC} the dynamics of the mode amplitude $a$ are now coupled to the conjugated mode amplitude $b^*$ and the EoMs in the signal frame reduce to

\begin{equation}
\begin{array}{r}    
\kappa_{a}\Delta_{a}a+\sqrt{\kappa_{a}\kappa_{b}}\beta_{ab}b^*=i\sqrt{\kappa_{a}^{\text{ext}}}a_{\text{in}},\\
\\
-\kappa_{b}\Delta_{b}^*b^*-\sqrt{\kappa_{a}\kappa_{b}}\beta_{ab}^*a=i\sqrt{\kappa_{b}^{\text{ext}}}b_{\text{in}}^*,
\end{array}  
\label{eq:EoM_PA}
\end{equation}

with $\Delta_{j}=(\omega_{j}^\text{s}-\omega_{j})/\kappa_{j}+i/2$ and $\beta_{ab}=|g_{ab}|e^{-i\phi_{ab}}/(2\sqrt{\kappa_a\kappa_b})$. Again, when comparing with the frequency conversion case, one can notice the sign change for the detuning term and the coupling term in the equation for $b^*$. These subtle differences lead to a very different scattering matrix $\mathbf{S}$, which in the ideal resonant case ($\omega_{ab}^p = \omega_{b} + \omega_{a}$ and $\omega_{j}^\text{s}=\omega_{j}$), and neglecting internal loss ($\kappa_{j}^\text{int}=0$), is

\begin{equation}
\mathbf{S}=\left(\begin{array}{cc}
\dfrac{1+4\abs{\beta_{ab}}^2}{1-4\abs{\beta_{ab}}^2} & \dfrac{4i\beta_{ab}}{1-4\abs{\beta_{ab}}^2} \\
\dfrac{-4i\beta_{ab}^*}{1-4\abs{\beta_{ab}}^2} & \dfrac{1+4\abs{\beta_{ab}}^2}{1-4\abs{\beta_{ab}}^2}
\end{array}\right)
\label{eq:S_PA_0}
\end{equation}

Close examination of the scattering matrix in Eq. \ref{eq:S_PA_0} reveals a divergence for $\abs{\beta_{ab}}=1/2$, in stark contrast with the frequency conversion case (Eq. \ref{eq:S_FC_0}). In the limit $\abs{\beta_{ab}}\rightarrow1/2^-$ each scattering parameter has an amplitude gain related to $\sqrt{G}\approx2/(1-4\abs{\beta_{ab}}^2)$, and the full signal flow is shown in Fig. \ref{figPA}(c). As in the frequency conversion case, note that $S_{ba}=S_{ab}^*$. In Fig. \ref{figPA}(d) we show the measured scattering parameters for various values of $\abs{\beta_{ab}}$, \textit{i.e.}, for various pump powers. As $\abs{\beta_{ab}}$ increases, so does the gain, at the expense of a typical reduction in linewidth \cite{heffner1958gain}. Good agreement is found with a fit to the solutions of Eq. \ref{eq:EoM_PA}.

The noise performance of the amplifier is shown in Fig. \ref{figPA}(e). In a separate experiment we calibrated an upper bound for the system added noise of the measurement setup at the reference plane of the FPJA, as explained in detail in Appendix C. When the FPJA is operated as an amplifier we measure a noise rise, and the knowledge of the noise floor in photon units allows us to extract the system added noise referred to the input of the FPJA. For $\abs{\beta_{ab}}=0$, we observe a slight degradation of the system added noise due to the internal loss of the resonators, up to $n_\text{add}\approx40~\text{photons}$ at $\omega_{a}$ and $n_\text{add}\approx30~\text{photons}$ at $\omega_{b}$. As the gain increases, the noise contribution of the measurement chain is overwhelmed and the system added noise decreases, down to $n_\text{add}\approx1.0\pm0.1~\text{photons}$ at $\omega_{a}$ and $\omega_{b}$, approaching the Standard Quantum Limit (SQL) of $n_\text{add}^\text{SQL}=0.5$. The system added noise plateaus at high gain, remaining slightly above the SQL. This could originate from excess thermal population of the resonators, potential excess loss in the FPJA packaging or simply from an offset in the reference plane of the noise calibration.

\subsubsection{The FPJA as a circulator}

\begin{figure*}
	\includegraphics[scale=1]{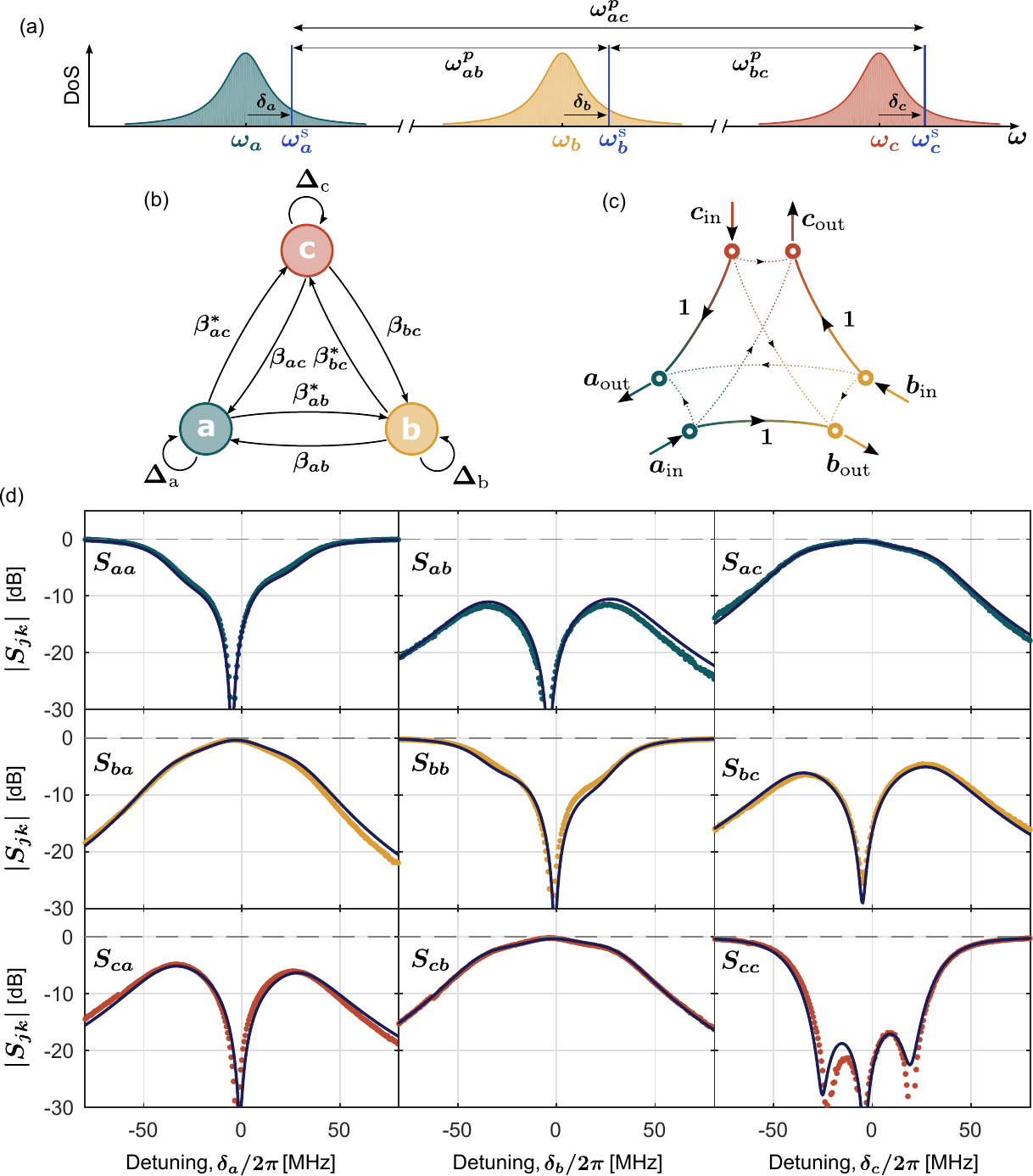}
	\caption{Circulator. (a) Frequency space diagram: Three pumps with frequencies $\omega_{ab}^p \approx \abs{\omega_{b}-\omega_{a}}$, $\omega_{bc}^p \approx \abs{\omega_{c}-\omega_{b}}$, $\omega_{ac}^p \approx \abs{\omega_{c}-\omega_{a}}$ and respective phase $\phi_{ab}$, $\phi_{bc}$ and $\phi_{ac}$ allow the conversion between signals at frequencies $\omega_{j}^\text{s} = \omega_{j}-\delta_j$ (${j} \in \{a,b,c\}$), closing a loop in frequency space. Interferences between paths enable nonreciprocal signal circulation. The circulation direction is set by the total loop phase, $\phi_\mathrm{loop}=\phi_{ab}+\phi_{bc}-\phi_{ac}$. (b) Graph representation of the EoM  \ref{eq:EoM_Circ}. (c) Ideal signal flow diagram. (d) Measured scattering parameters (dots) and fits to Eq. \ref{eq:EoM_Circ} (lines) as a function of the detuning $\delta_j$, for fixed pump powers ($\abs{\beta_{jk}}=1/2$) and loop phase ($\phi_\mathrm{loop}\approx-\pi/2$).
		\label{figCirculator}}
\end{figure*}

By connecting all three resonators via frequency conversion, we can build the first nontrivial mode of operation: the circulator. Using three pumps, we modulate the flux through the SQUID at the difference frequencies $\omega_{jk}\approx\abs{\omega_{j}-\omega_{k}}$, where ${j},{k} \in \{{a,b,c}\}$, satisfying the condition $\omega_{ab}^p + \omega_{bc}^p = \omega_{ac}^p$ (see Fig. \ref{figCirculator}(a)). Effectively, this forms a closed loop in frequency space connecting input and output signals at $\omega_{j}^\text{s}$, where ${j} \in \{{a,b,c}\}$. The EoMs in the signal frame reduce to:

\begin{equation}
\begin{array}{l}    
\kappa_{a}\Delta_{a}a +\sqrt{\kappa_{a}\kappa_{b}}\beta_{ab}b +\sqrt{\kappa_{a}\kappa_{c}}\beta_{ac}^*c =i\sqrt{\kappa_{a}^{{ext}}}a_{\text{in}}, \\    
\\
\kappa_{b}\Delta_{b}b+\sqrt{\kappa_{b}\kappa_{c}}\beta_{bc}c + \sqrt{\kappa_{a}\kappa_{b}}\beta_{ab}^*a =i\sqrt{\kappa_{b}^{\text{ext}}}b_{\text{in}}, \\
\\
\kappa_{c}\Delta_{c}c+\sqrt{\kappa_{a}\kappa_{c}}\beta_{ac}a+\sqrt{\kappa_{b}\kappa_{c}}\beta_{bc}^*b =i\sqrt{\kappa_{c}^{\text{ext}}}c_{\text{in}},
\end{array}  
\label{eq:EoM_Circ}
\end{equation}

with $\Delta_j=(\omega_{j}^\text{s}-\omega_{j})/\kappa_j+i/2$ and $\beta_{jk}=|g_{jk}|e^{i\phi_{jk}}/(2\sqrt{\kappa_j\kappa_k})$, where ${j},{k} \in \{{a,b,c}\}$. The graph representation of this equation is shown in Fig. \ref{figCirculator}(b). This loop topology is at the heart of the nonreciprocal behavior of this mode of operation. Indeed, by closing this loop we build an interferometer, where the phase shift in each arm is direction dependent. The interference is controlled by the loop phase $\phi_\mathrm{loop}=\phi_{ab}+\phi_{bc}-\phi_{ac}$, where $\phi_{jk}$ is the phase of the pump connecting modes ${j}$ and ${k}$. In the ideal resonant case $\omega_{jk} = \abs{\omega_{j} - \omega_{k}}$ and $\omega_{j}^\text{s}=\omega_{j}$, neglecting internal loss, tuning each coupling strength to produce ideal frequency conversion ($\abs{\beta_{jk}}=1/2$), and for $\phi_\mathrm{loop}=-\pi/2$, the scattering matrix $\mathbf{S}$ for the system is

\begin{equation}
\mathbf{S}=\left(\begin{array}{ccc}
0 & 0 & 1 \\
1 & 0 & 0 \\
0 & 1 & 0 
\end{array}\right)
\label{eq:S_Circulator_0}
\end{equation}

In this case the system is matched, with $\abs{S_{jj}}=0$, and nonreciprocal, $\abs{S_{jk}}\neq\abs{S_{kj}}$, corresponding to the signal flow diagram in Fig. \ref{figCirculator}(c). An input signal in mode $a$ circulates to mode $b$ then $c$. The direction of circulation is controlled by the loop phase and can be reversed by setting $\phi_\mathrm{loop}=\pi/2$. 

Experimentally we start by tuning the pumps to produce ideal frequency conversion between each pair of modes separately, and then simultaneously turn on all three pumps. The measured scattering parameters for $\phi_\mathrm{loop}=-\pi/2$ are shown in Fig. \ref{figCirculator}(d). The full dependence with the loop phase $\phi_\mathrm{loop}$ is shown in Fig. \ref{figSIcirculator}. Very good agreement is obtained with solutions of Eq. \ref{eq:EoM_Circ}. With return loss exceeding 20~dB the device exhibits excellent impedance match at all three modes. We measure a transmission efficiency of more than $-0.5~\mathrm{dB}$ and an isolation exceeding  $20~\text{dB}$ over a $6~\text{MHz}$ bandwidth. A transmission efficiency of more than $-1~\mathrm{dB}$ and an isolation exceeding  $10~\text{dB}$ is maintained over a $60~\text{MHz}$ bandwidth.

\subsubsection{The FPJA as a directional amplifier}

\begin{figure*}
	\includegraphics[scale=1]{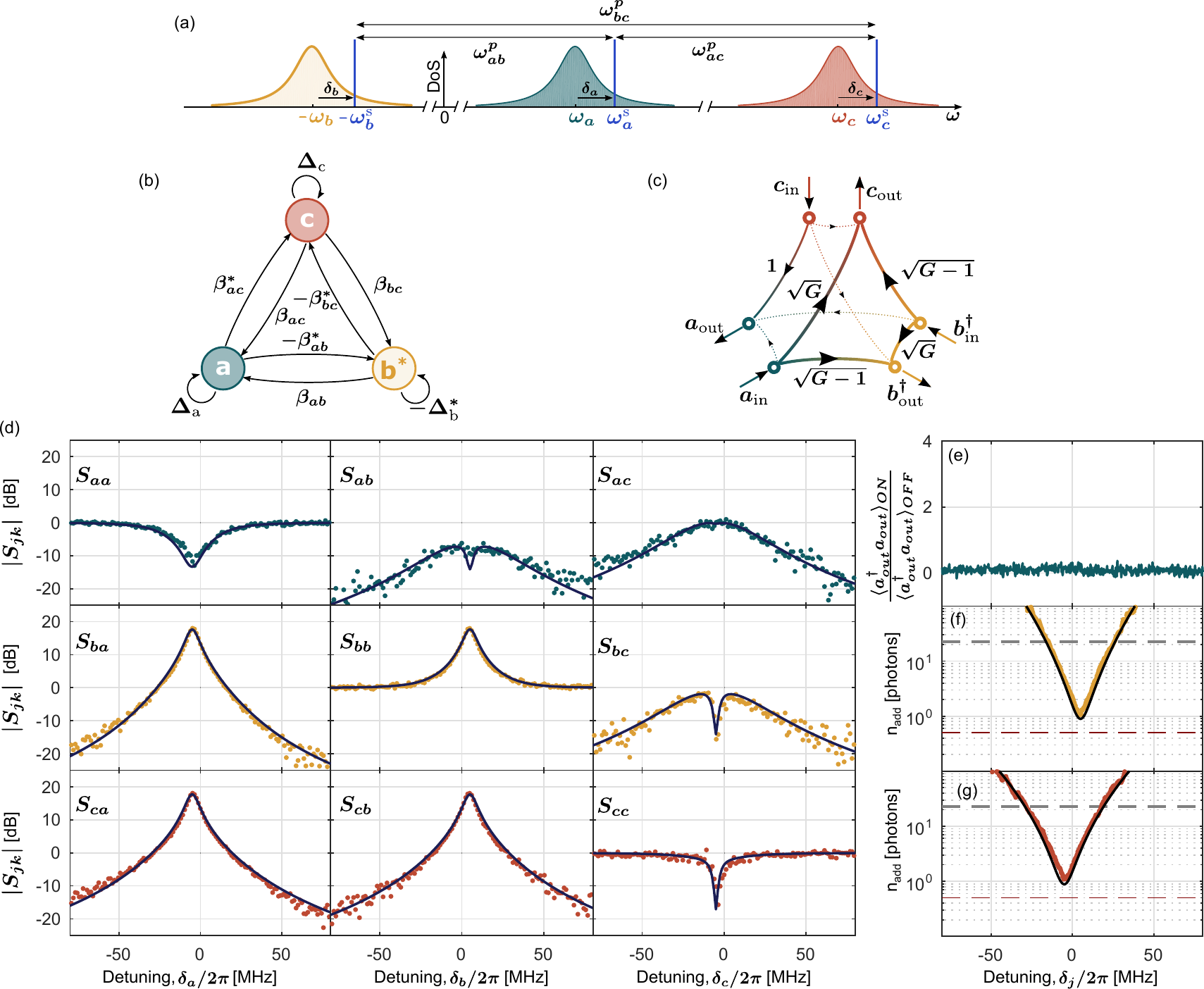}
	\caption{Directional phase-preserving amplifier. (a) Frequency space diagram: Three pumps with frequencies $\omega_{ac}^p\approx\abs{\omega_{c}-\omega_{a}}$, $\omega_{ab}^p\approx\omega_{b}+\omega_{a}$ and $\omega_{bc}^p\approx\omega_{c}+\omega_{b}$ and respective phase $\phi_{ab}$, $\phi_{bc}$ and $\phi_{ac}$ allow the conversion between signals at the frequencies $\omega_{a}^\text{s}$ , $-\omega_{b}^\text{s}$ and $\omega_{c}^\text{s}$, closing a loop in frequency space. Interferences between paths enable nonreciprocal signal amplification. The amplification direction is set by the total loop phase, $\phi_\mathrm{loop}=\phi_{ab}+\phi_{bc}+\phi_{ac}$.  (b) Graph representation of the EoM \ref{eq:EoM_Delta}. (c) Ideal signal flow diagram. (d) Measured scattering parameters (dots) and fits to Eq. \ref{eq:EoM_Delta} (lines) as a function of the detuning $\delta_j$, for fixed pump powers ($\abs{\beta_{jk}}\approx1/2$) and loop phase ($\phi_\mathrm{loop}\approx-\pi/2$). (e) Return noise of the amplifier: ratio of the measured noise power out of mode $a$ for the pumps on and off, showing that no extraneous noise is added by the amplifier. (f) and (g) Measured system added noise (dots) and theoretical predictions (lines) referred to the input of the FPJA, as a function of the detuning $\delta_j$, respectively for mode $b$ and $c$.
		\label{figDelta}}
\end{figure*}

When operated as a two-mode amplifier the FPJA is non-directional (see Fig. \ref*{figPA}), and requires a circulator to separate the input and output signals. In this section we circumvent this requirement by operating the FPJA as a directional amplifier.  Using three pumps, we modulate the flux through the SQUID at frequencies $\omega_{ac}^p\approx\omega_{c}-\omega_{a}$, $\omega_{ab}^p\approx\omega_{b}+\omega_{a}$ and $\omega_{bc}^p\approx\omega_{c}+\omega_{b}$, satisfying the overall loop closure condition $\omega_{ac}^p + \omega_{ab}^p = \omega_{bc}^p$ (see Fig. \ref{figDelta}(a)). This corresponds to simultaneously connecting modes $a$ and $c$ via frequency conversion, while connecting modes $a$ and $b$ and modes $b$ and $c$ via amplification, forming a loop in frequency space. The resulting EoMs are

\begin{equation}
\begin{array}{r}    
\kappa_{a}\Delta_{a}a+\sqrt{\kappa_{a}\kappa_{b}}\beta_{ab}b^* +\sqrt{\kappa_{a}\kappa_{c}}\beta_{ac}c=i\sqrt{\kappa_{a}^{\text{ext}}}a_{\text{in}} \\
\\
-\kappa_{b}\Delta_{b}^*b^*+\sqrt{\kappa_{b}\kappa_{c}}\beta_{bc}c -\sqrt{\kappa_{a}\kappa_{b}}\beta_{ab}^*a =i\sqrt{\kappa_{b}^{\text{ext}}}b_{\text{in}}^* \\
\\
\kappa_{c}\Delta_{c}c+\sqrt{\kappa_{a}\kappa_{c}}\beta_{ac}^*a-\sqrt{\kappa_{b}\kappa_{c}}\beta_{bc}^*b^*=i\sqrt{\kappa_{c}^{\text{ext}}}c_{\text{in}}
\end{array} 
\label{eq:EoM_Delta}
\end{equation}

The graph representation of these EoMs is shown in Fig. \ref{figDelta}(b). Similarly to the circulator case, the loop phase $\phi_\mathrm{loop}=\phi_{ab}+\phi_{bc}+\phi_{ac}$ controls the direction of the amplification. In the ideal resonant case, neglecting internal loss, tuning the coupling strength to produce ideal frequency conversion between modes $a$ and $c$, $\abs{\beta_{jk}}=1/2$, for symmetric amplification coupling strength $\abs{\beta_{ab}}=\abs{\beta_{bc}}$,  and for $\phi_\mathrm{loop}=-\pi/2$, the scattering matrix $\mathbf{S}$ is

\begin{equation}
\mathbf{S}=\left(\begin{array}{ccc}
0 & 0 & 1 \\
\sqrt{G-1} & \sqrt{G} & 0 \\
\sqrt{G} & \sqrt{G-1}& 0 
\end{array}\right)
\label{eq:S_Delta_0}
\end{equation}

where $\sqrt{G} = \left(1+4\abs{\beta_{ab}}^2\right)/\left(1-4\abs{\beta_{ab}}^2\right)$. In this configuration, mode $a$ serves as the input port and is impedance matched (no reflection). An input signal is amplified toward both output modes $b$ and $c$. The amplified signal at each output is added to the amplified vacuum seeded into mode $b$, resulting in the same minimum system added noise as for a standard two-mode amplifier. Finally, vacuum noise seeded into mode $c$ is routed to mode $a$ with unity gain. The measured scattering parameters for $\phi_\mathrm{loop}=\pi/2$ are shown in Fig. \ref{figDelta}(d). The full dependence with the loop phase $\phi_\mathrm{loop}$ is shown in Fig. \ref{figSIdelta}. Good agreement is obtained with numerical solutions of Eq. \ref{eq:EoM_Delta}. We obtain a forward gain of $18~\text{dB}$ while maintaining good impedance matching (return loss in excess of $10~\text{dB}$) and isolation (in excess of $8~\text{dB}$). The noise performances are shown in Fig. \ref{figDelta} (e), (f) and (g), corresponding to the respective output of mode $a$, $b$ and $c$. The system added noise measured at the output of mode $b$ and $c$, referred to the input of mode $a$, approaches the quantum limit with $n_\text{add}\approx1.1\pm0.1~\text{photons}$. Moreover, due to the directionality of the amplifier, we do not observe any noise rise at the output of mode $a$, and therefore do not expect any backaction on a future DUT ---a crucial property for the integrated measurement of a microwave quantum system.

\section{Discussion and Conclusion}
We have demonstrated the ability to program \textit{in situ} a low-loss superconducting device to perform reciprocal and nonreciprocal analog microwave signal processing close to the quantum limit. This technology is compatible with most superconducting quantum computation systems, opening the way to full integrability and scalability.
Finally, we emphasize that we have shown here only a small subset of all the possible networks of parametrically coupled modes. Other networks, using three or more modes, could for eample lead to phase sensitive directional amplification, or alleviate the gain-bandwidth product. We note that it would be straightforward to design a circuit with bare resonance frequencies placed between $2~\mathrm{GHz}$ and $20~\mathrm{GHz}$ and with linewidths individually set between $5~\mathrm{GHz}$ and $500~\mathrm{MHz}$. Finally, routing the different mode frequencies to separate physical ports relies simply on the design of a low-loss on-chip microwave multiplexer, directly compatible with the device fabrication.

\section*{Acknowledgment}
This work was supported by the NIST Quantum Information Program. This article is a contribution of the U.S. government, not subject to copyright.

\section*{Appendix A: Device fabrication and layout}

\begin{figure*}
	\includegraphics[scale=1]{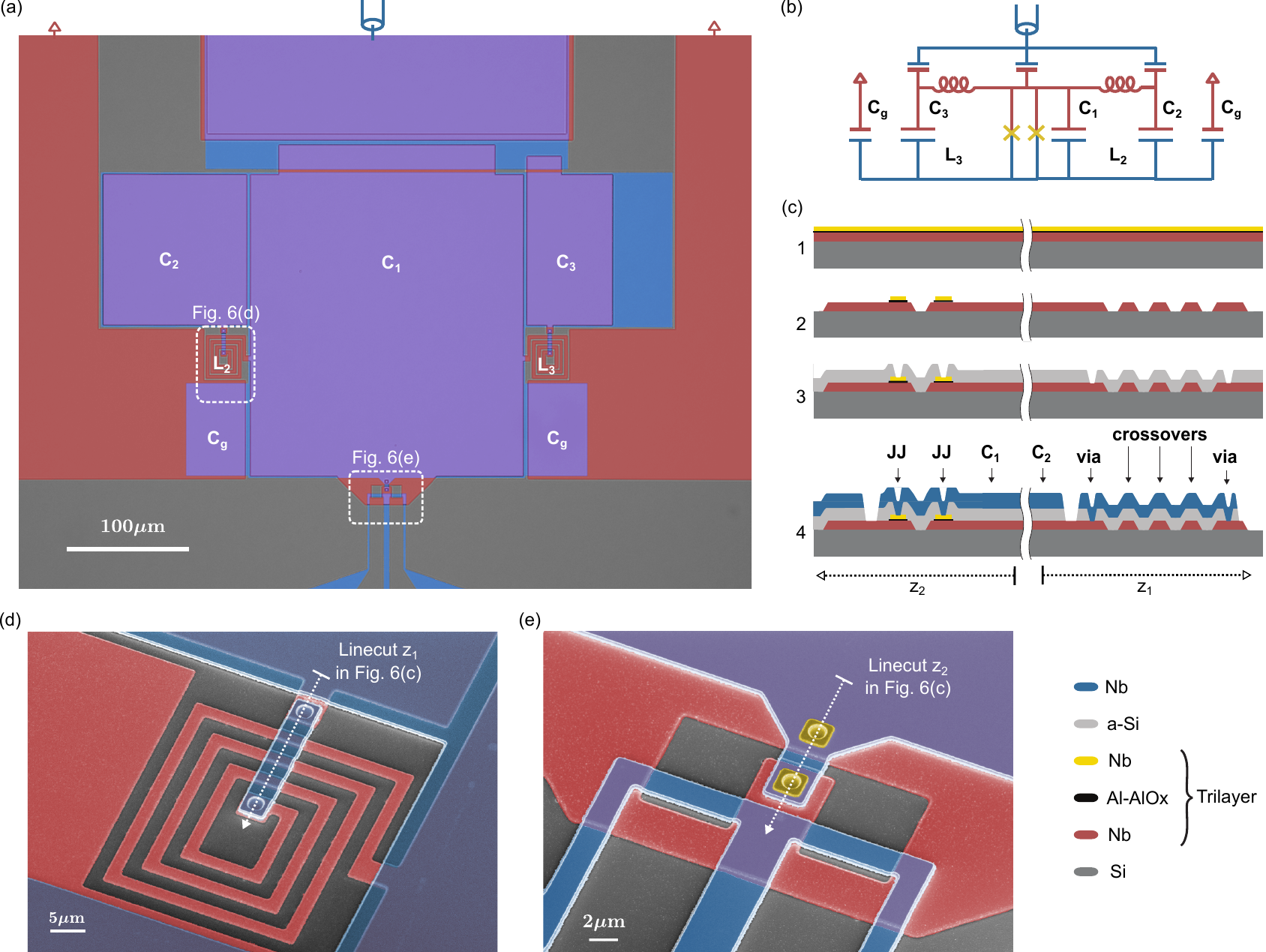}
	\caption{Device fabrication and layout. (a) False colored optical micrograph. The silicon substrate is in grey, the bottom $\mathrm{Nb}$ layer in red and the top $\mathrm{Nb}$ layer in blue. Overlap between the two $\mathrm{Nb}$ layers forms parallel plates capacitors, in purple. (b) Circuit equivalent of the device. (c) Fabrication process, along the cross-sections $z_1$ and $z_2$ shown in Fig. (d) and (e). Step 1, the Nb/Al-AlO$_x$/Nb trilayer is prepared on intrinsic silicon wafer. Step 2, the trilayer is patterned top-down in three steps to define Josephson junctions and the bottom Nb wiring layer.  Step 3, the a-Si dielectric layer is deposited and patterned to define vias. Step 4, the Nb layer is deposited over a-Si and vias, and is patterned to define the top wiring layer.  Any uncovered a-Si is also removed in this step.  (d) and (e) Scanning electron micrograph of one of the coil inductors, and of the gradiometric SQUID, using the same color scheme, with additionally the Josephson junction in yellow.
		\label{figSIdevice}}
\end{figure*}

The device is fabricated with optical lithography by using a Nb/Al-AlO$_x$/Nb trilayer process to form Josephson junctions, and by utilizing amorphous silicon (a-Si) as a low-loss interlayer dielectric.   The circuit layout and pictures are shown in Fig. \ref{figSIdevice}. The device fabrication is summarized in Fig. \ref{figSIdevice}(c):

\begin{itemize}
	\item A Nb/Al-AlO$_x$/Nb trilayer is prepared on a high-resistivity intrinsic silicon wafer  ($>20~\mathrm{k\Omega\cdot cm}$) by subsequently sputtering a $200~\text{nm}$ $\mathrm{Nb}$ layer (red), a $8~\text{nm}$ $\mathrm{Al}$ layer which is then oxidized to form a tunnel barrier (black), and finally  a $110~\text{nm}$ $\mathrm{Nb}$ (yellow).  The preparation of the trilayer is performed \textit{in situ} in a sputtering deposition tool without breaking vacuum. 
	\item The trilayer is patterned top-down in three iterations of optical lithography followed by material etching. First, the Josephson junctions (JJ) areas are defined by etching the top Nb layer (yellow) using a vertical plasma etch (SF$_6$/O$_2$). Second, the excess Al-AlOx (black) is removed away from the junctions by a wet etch (MF-26A, which is also the optical resist developer).  Finally, we define the bottom wiring layer in Nb (red) using a sloped plasma etch (CF$_4$/O$_2$).  
	\item The Josephson junctions and bottom wiring layer are covered by a $300~\text{nm}$ amorphous silicon layer (a-Si) deposited by plasma enhanced chemical vapor deposition (PECVD). Vias are then defined by etching into the a-Si layer using a sloped plasma etch (SF$_6$/O$_2$).
	\item After an in situ RF clean, a $300~\text{nm}$ $\mathrm{Nb}$ layer is sputtered and patterned by a vertical plasma etch (SF$_6$) to define the top wiring layer (blue). This etch also removes any uncovered a-Si.
\end{itemize}

The properties of the a-Si were characterized by independent measurements of lumped-element LC resonators fabricated on the same wafer than the FPJA, yielding a relatively high dielectric constant ($\epsilon_r\approx9$) and low loss tangent ($1.5$---$5\times10^{-4}$). This allows for the design of compact and low-loss lumped-element capacitor and inductors. The Josephson junctions have an area of $2.5~\mathrm{\mu m}\times2.5~\mathrm{\mu m}$, and the oxidation parameters lead to a critical current $I_c=5~\mathrm{\mu A}$ per junction. The SQUID loop has a gradiometric design, first-order insensitive to magnetic fields.

The design of the device is based on a circuit with three discrete poles, each tunable by a single SQUID inductance. This is achieved here by connecting three LC resonators in parallel ($L_{\mathrm{SQ}}C_1$, $L_{2}C_2$ and $L_{3}C_3$), as shown in Fig. \ref*{figSIdevice}. Alternatively this device can be understood as two LC resonators, $L_{2}C_2$ and $L_{3}C_3$, coupled via a resonant coupling element (formed by the SQUID shunted by $C_1$). The resonances of this circuit, denoted $a$, $b$ and $c$, tune with the SQUID flux, as shown in Fig. \ref{figSetup}. Throughout this work we use the SQUID as a tunable linear coupling element, neglecting the intrinsic non-linearity of the SQUID. To understand the origin of the bilinear coupling between the three modes, one can consider the energy stored in the SQUID, $E_{\mathrm{SQ}}\propto L_{\mathrm{SQ}}I_{\mathrm{SQ}}^2$. For each mode $j$, a fraction $\alpha_j$ of the mode current $I_j$ flows through the SQUID such that $I_{\mathrm{SQ}}=\sum_{j}\alpha_{j}I_j$. This results in $E_{\mathrm{SQ}}\propto\sum_{j,k}\alpha_j\alpha_k L_{\mathrm{SQ}}I_jI_k$, effectively producing linear coupling of modes $j$ and $k$.

\section*{Appendix B: Theory of parametrically coupled modes}

\begin{figure*}
	\includegraphics[scale=1]{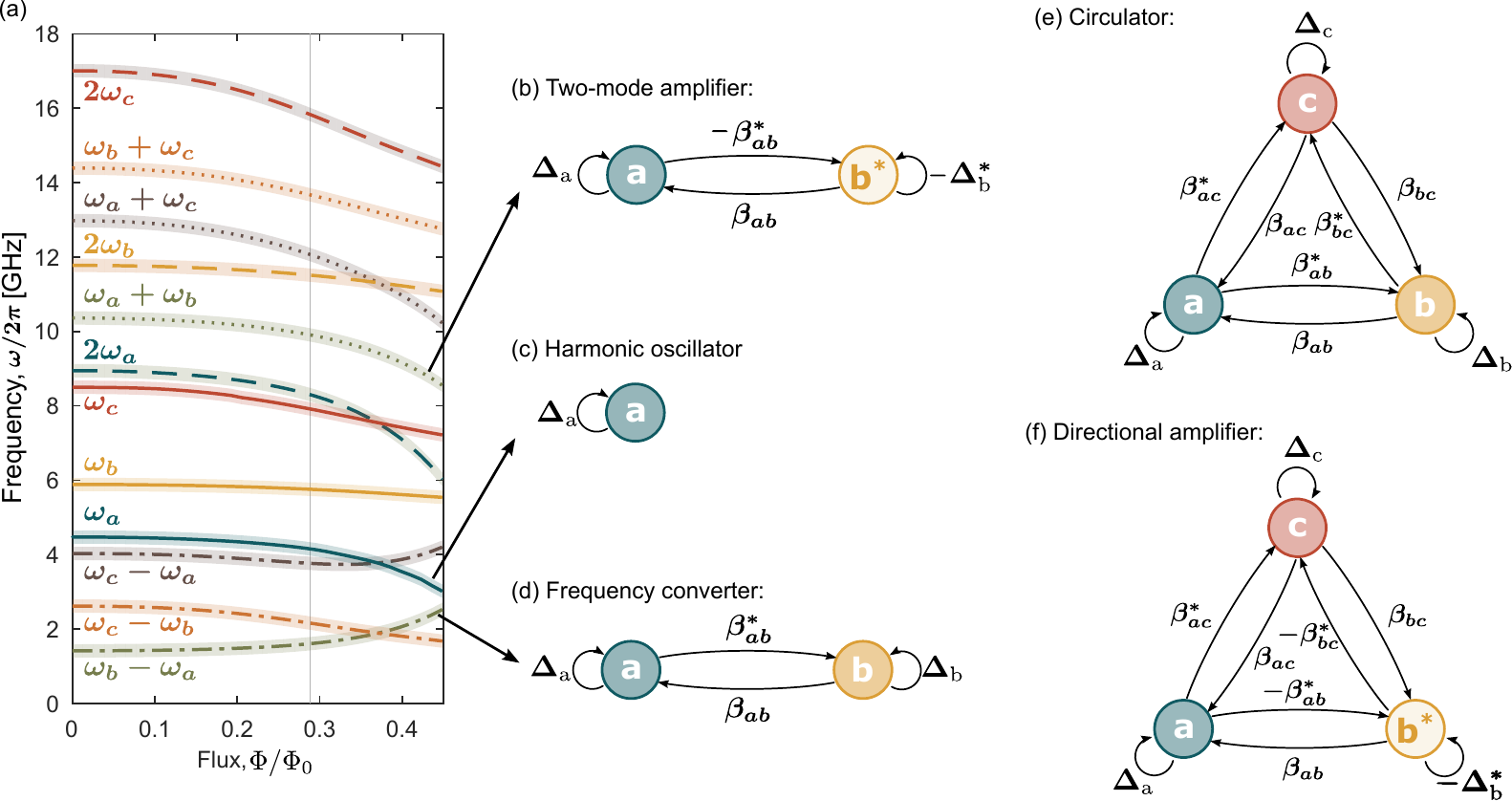}
	\caption{Process spectroscopy. (a) Mode frequencies $\omega_{j}$, and first-order modulation frequencies $\abs{\omega_{j}\pm\omega_{k}}$ (${j},{k} \in \{{a,b,c}\}$) as a function of flux. The shaded areas represent a bandwidth of $180~\text{MHz}$, necessary to ensure a good rotating wave approximation. Graph representation of the EoM for (b) a single harmonic oscillator (c) frequency conversion between two modes, (d) parametric amplification between two modes, (e) circulation and (f) directional phase-preserving amplification.
		\label{figSItheory}}
\end{figure*}

In this section we describe our approach to solving the equations of motion (EoMs) for a system of parametrically coupled modes. We aim at deriving the resulting scattering parameters (section B.1 to B.4) and output noise (section B.5). We begin with the case of a single damped and driven oscillator to introduce the concepts and our notation. We then describe the building blocks of parametric physics, namely frequency conversion and amplification between two modes. Combining these processes in three-mode systems allows us to describe the circulator and directional amplifier. We largely reproduce concepts and style from Ranzani and Aumentado \cite{Ranzani2015}, with minor changes of notation and normalization.

\subsubsection{A single driven and damped oscillator}

The Hamiltonian for a driven harmonic oscillator with resonant frequency $\omega_{a}$ and loss rate $\kappa_a$, after tracing over bath modes in the rotating wave approximation, is
\begin{equation}
\frac{\hat{\mathcal{H}}}{\hbar}=\left(\omega_{a}-i\frac{\kappa_a}{2}\right)\hat{a}^{\dagger}\hat{a}+i\sqrt{\kappa_a^{\text{ext}}}\left(\hat{a}_{\text{in}}-\hat{a}_{\text{in}}^{\dagger}\right)\left(\hat{a}+\hat{a}^{\dagger}\right), \label{eq:H_ho}
\end{equation}
where $\kappa_a^{\text{ext}}$ is the external coupling to the drive port, and $\hat{a}$ and $\hat{a}_{\text{in}}$ are the time-dependent annihilation operators for the internal mode and input drive respectively. To simplify the notation we choose (1) to include the loss as an imaginary component of the resonant frequency, making the Hamiltonian non-Hermitian and (2) to have the phase quadrature of the drive coupled to the amplitude quadrature of the internal mode. Note that to preserve quantum commutators, Eq. \ref{eq:H_ho} needs an extra noise input term proportional to $\sqrt{\kappa_a^{\text{int}}}$ so that $\kappa_a=\kappa_a^{\text{ext}}+\kappa_a^{\text{int}}$ (see section B.4).

The Heisenberg-Langevin EoM is $\dot{\hat{a}} = -i\left[\hat{a},\hat{\mathcal{H}}/\hbar\right]$. To analyze the EoM, we consider an input signal at the signal frequency $\omega_a^\text{s}$ and move to a rotating frame at that frequency by defining new annihilation operators $\hat{a}\rightarrow \hat{a}\exp(-i\omega_{a}^\text{s}t)$, and $\hat{a}_{\text{in}}\rightarrow \hat{a}_{\text{in}}\exp(-i\omega_{a}^\text{s}t)$, so that $\hat{a}$ and $\hat{a}_{\text{in}}$ are time-independent in this frame. Additionally, we will study the dynamics of the expectation values, defining $a \equiv \left< \hat{a}\right>$. The equation of motion reads
\begin{equation}
\kappa_a\Delta_aa=i\sqrt{\kappa_a^{\text{ext}}}a_{\text{in}}, \label{eq:eom_ho}
\end{equation}
where
\begin{equation}
\Delta_a=\frac{\omega_{a}^\text{s}-\omega_{a}}{\kappa_a}+\frac{i}{2},
\end{equation}
is a normalized complex detuning parameter. In the above we have applied the rotating wave approximation (RWA), neglecting a term with explicit time-dependence in the signal frame. We assume that the frequency of this term, $2\omega_a^\text{s}$, is much larger than the relevant linewidth of the oscillator, $\kappa_a$, so that the susceptibility, $\chi=i/(\kappa_a\Delta_a)$, at those frequencies is small enough to not cause any appreciable dynamics. 

To introduce the graph representation that we will use in the following cases, Fig.~\ref{figSItheory}(c) shows the graph corresponding to the single-oscillator EoM~\eqref{eq:eom_ho}. It consists of a single vertex for the mode amplitude $a$, with a \textit{self-loop} associated with the complex detuning $\Delta_a$.

The output field at the signal frequency $\omega_{a}^\text{s}$ is obtained via the input-output relation $a_{\text{out}}=\sqrt{\kappa_a^{\text{ext}}}a-a_{\text{in}}$, finally leading to the familiar Lorentzian form of the reflection coefficient:
\begin{equation}
S_{aa} = \frac{a_{\text{out}}}{a_{\text{in}}} = \left(\frac{\kappa_a^{\text{ext}}}{\kappa_a}\frac{i}{\Delta_{a}}-1\right).
\end{equation}

\subsubsection{Two coupled modes}

Let us now consider two oscillators with resonant frequencies $\omega_{a}$ and $\omega_{b}$, loss rates $\kappa_a$ and $\kappa_b$, and a time-dependent coupling rate $g_{ab}(t)=\abs{g_{ab}}\cos(\omega_{ab}^pt+\phi_{ab})$. The Hamiltonian of the system is
\begin{multline}
\hat{\mathcal{H}}/\hbar=\left(\omega_{a}-i\frac{\kappa_a}{2}\right)\hat{a}^{\dagger}\hat{a} +i\sqrt{\kappa_a^{\text{ext}}}\left(\hat{a}_{\text{in}}-\hat{a}_{\text{in}}^{\dagger}\right)\left(\hat{a}+\hat{a}^{\dagger}\right) \\
+ \left(\omega_{b}-i\frac{\kappa_b}{2}\right)\hat{b}^{\dagger}\hat{b} +i\sqrt{\kappa_b^{\text{ext}}}\left(\hat{b}_{\text{in}}-\hat{b}_{\text{in}}^{\dagger}\right)\left(\hat{b}+\hat{b}^{\dagger}\right) \\
- g_{ab}(t)\left(\hat{a}+\hat{a}^{\dagger}\right)\left(\hat{b}+\hat{b}^{\dagger}\right), \label{eq:H_2modes}
\end{multline}
where $\kappa_a^{\text{ext}}$ and  $\kappa_b^{\text{ext}}$ are the coupling rates to the external port, and $\hat{a}$, $\hat{b}$, $\hat{a}_{\text{in}}$ and $\hat{b}_{\text{in}}$ are the time-dependent annihilation operators for the internal mode and input drives for each oscillator, respectively. We consider the oscillators to be driven at the frequencies $\omega_a^\text{s}$ and $\omega_b^\text{s}$, related to each other by the pump frequency of $\omega_{ab}^p$. We move to a co-rotating frame, and define new (time-independent) annihilation operators $\hat{a}\rightarrow \hat{a}\exp(-i\omega_{a}^\text{s}t)$, $\hat{b}\rightarrow \hat{b}\exp(-i\omega_{b}^\text{s}t)$, $\hat{a}_{\text{in}}\rightarrow \hat{a}_{\text{in}}\exp(-i\omega_{a}^\text{s}t)$, and $\hat{b}_{\text{in}}\rightarrow \hat{b}_{\text{in}}\exp(-i\omega_{b}^\text{s}t)$. 

In the case of resonant coupling, $g_{ab}$ is time-independent, so $\omega_a^\text{s}=\omega_b^\text{s}$. In this case, if the two oscillators are detuned from one another ($\abs{\omega_b - \omega_a} \gg \kappa_a,\kappa_b$), then the large detuning (small susceptibility) prevents appreciable energy transfer between the oscillators. If the pump frequency is near the sum or difference of the two resonant frequencies, however, then both detunings can be small (with corresponding large susceptibility) to allow energy transfer between the two oscillators. When $\omega_{ab}^p\approx\abs{\omega_b-\omega_a}$, a pump photon can bridge the energy gap between the two oscillators allowing for frequency conversion. When $\omega_{ab}^p\approx\omega_a+\omega_b$, a pump photon can be down-converted into a photon in each oscillator, amplifying each oscillator's amplitude.

\paragraph{Frequency conversion}

Consider the case where the coupling term is modulated near the difference frequency $\omega_{ab}^p\approx\abs{\omega_b-\omega_a}$. To simplify notation, let us choose $\omega_b>\omega_a$, such that $\omega_b^\text{s}=\omega_a^\text{s}+\omega_{ab}^p$. The EoMs after the RWA become
\begin{equation}
\begin{aligned}
\kappa_a\Delta_aa+\sqrt{\kappa_a\kappa_b}\beta_{ab}b=i\sqrt{\kappa_a^{\text{ext}}}a_{\text{in}} ,  \\
\kappa_b\Delta_bb+\sqrt{\kappa_a\kappa_b}\beta_{ab}^*a=i\sqrt{\kappa_b^{\text{ext}}}b_{\text{in}},
\end{aligned} \label{eq:eom_fc}
\end{equation}
where $\Delta_{j}=(\omega_{j}^\text{s}-\omega_{j})/\kappa_{j}+i/2$ are normalized complex detunings, and  $\beta_{ab}=|g_{ab}| e^{i\phi_{ab}}/(2\sqrt{\kappa_a\kappa_b})$ is the normalized complex coupling between the mode amplitudes $a$ and $b$.

When working with multiple modes it becomes useful to adopt a matrix representation of these equations. We define a vector of intra-cavity mode amplitudes $\mathbf{A}=(a,b)^T$, input mode amplitudes $\mathbf{A_{\text{in}}}=(a_{\text{in}},b_{\text{in}})^T$, output mode amplitudes $\mathbf{A_{\text{out}}}=(a_{\text{out}},b_{\text{out}})^T$, diagonal matrices for the total loss rates $\mathbf{K}=\text{diag}(\sqrt{\kappa_a},\sqrt{\kappa_b})$, and external couplings $\mathbf{K^{\text{ext}}}=\text{diag}(\sqrt{\kappa_a^{\text{ext}}},\sqrt{\kappa_b^{\text{ext}}})$, and finally the mode-coupling matrix $\mathbf{M}$ so that Eq.~\eqref{eq:eom_fc} become $\mathbf{K}\mathbf{M}\mathbf{K}\mathbf{A} = i\mathbf{K^{\text{ext}}}\mathbf{A_{\text{in}}}$, where
\begin{equation}
\mathbf{M}=\begin{pmatrix}\Delta_{a} & \beta_{ab}\\
\beta_{ab}^{*} & \Delta_{b}
\end{pmatrix}.
\end{equation}

The graph representation of the mode-coupling matrix $\mathbf{M}$ is shown in Fig.~\ref{figSItheory}(d). Each vertex corresponds to a mode amplitude, with self-loops associated with the complex detuning $\Delta_{j}$. The vertices are connected by coupling edges associated with the coupling terms $\beta_{ab}$ and $\beta_{ab}^*$.

We can solve these EoMs to calculate the scattering matrix $\mathbf{S}=\mathbf{A_{\text{out}}}^T/\mathbf{A_{\text{in}}}$, resulting in
\begin{equation}
\mathbf{S} = i\mathbf{H}\mathbf{M}^{-1}\mathbf{H}-\mathds{1}, \label{eq:eom_S}
\end{equation}
where we introduce the matrix $\mathbf{H}=\text{diag}\left( \sqrt{\eta_a},\sqrt{\eta_b}\right) $, where  $\eta_{j}=\kappa_{j}^{\text{ext}}/\kappa_{j}$ are coupling efficiency parameters characterizing the degree to which each mode is overcoupled. Expanding Eq.~\eqref{eq:eom_S} results in
\begin{equation}
\mathbf{S}=\left(\begin{array}{ll}
\dfrac{i\eta_a\Delta_b}{\Delta_a\Delta_b-\abs{\beta_{ab}}^2}-1 & \dfrac{-i\sqrt{\eta_a\eta_b}\beta_{ab}}{\Delta_a\Delta_b-\abs{\beta_{ab}}^2} \\
\dfrac{-i\sqrt{\eta_a\eta_b}\beta^*_{ab}}{\Delta_a\Delta_b-\abs{\beta_{ab}}^2} & \dfrac{i\eta_b\Delta_a}{\Delta_a\Delta_b-\abs{\beta_{ab}}^2}-1
\end{array}\right).
\label{eq:S_FC}
\end{equation}
On resonance ($\Delta_{j} = i/2$) and neglecting internal loss ($\eta_{j}=1$) one recovers Eq.~\eqref{eq:S_FC_0}:
\begin{equation*}
\mathbf{S}=\left(\begin{array}{cc}
\dfrac{1-4\abs{\beta_{ab}}^2}{1+4\abs{\beta_{ab}}^2} & \dfrac{4i\beta_{ab}}{1+4\abs{\beta_{ab}}^2} \\
\dfrac{4i\beta_{ab}^*}{1+4\abs{\beta_{ab}}^2} & \dfrac{1-4\abs{\beta_{ab}}^2}{1+4\abs{\beta_{ab}}^2}
\end{array}\right).
\end{equation*}
Unity transmission coincides with impedance matching (zero reflection) for $\abs{\beta_{ab}}=1/2$. Note that when $\beta_{ab}$ is complex, $S_{ab} \neq S_{ba}$ (although $|S_{ab}| = |S_{ba}|$). However, for modes at different frequencies, the phase shift between $S_{ab}$ and $S_{ba}$ must be measured relative to some reference whose phase is arbitrary. In a two-mode system, there can always be found a reference phase such that $\beta_{ab}$ is real and $\mathbf{S}$ is reciprocal. Yet, the phase dependence of $\mathbf{S}$ will become crucial when discussing three-mode systems, in which the phase reference is built into the system. Finally, note that for resonant coupling, where $\beta_{ab}$ is required to be real, the scattering matrix is unambiguously reciprocal.

\paragraph{Parametric amplification}

We now consider the case where the coupling is modulated near the sum frequency $\omega_{ab}^p\approx\omega_a+\omega_b$, such that $\omega^\text{s}_b=\omega_{ab}^p - \omega^\text{s}_a$. The EoMs after applying the RWA become
\begin{equation}
\begin{aligned}
\kappa_a\Delta_aa\medspace+\sqrt{\kappa_a\kappa_b}\beta_{ab}b^{*}=i\sqrt{\kappa_a^{\text{ext}}}a_{\text{in}} , \\
-\kappa_b\Delta_b^*b^{*}-\sqrt{\kappa_a\kappa_b}\beta_{ab}^*a=i\sqrt{\kappa_b^{\text{ext}}}b_{\text{in}}^{*},
\end{aligned} \label{eq:eom_pa}
\end{equation}
or, in matrix form $\mathbf{K}\mathbf{M}\mathbf{K}\mathbf{A} = i\mathbf{K^{\text{ext}}}\mathbf{A_{\text{in}}}$, where
\begin{equation}
\mathbf{M}=\begin{pmatrix}\Delta_{a} & \beta_{ab}\\
-\beta_{ab}^{*} & -\Delta_{b}^*
\end{pmatrix},
\end{equation}
and $\mathbf{A}=(a,b^{*})^T$, $\mathbf{A_{\text{in}}}=(a_{\text{in}},b_{\text{in}}^{*})^T$,  $\mathbf{A_{\text{out}}}=(a_{\text{out}},b_{\text{out}}^{*})^T$. Here note that for parametric amplification $\beta_{ab}=|g_{ab}|e^{-i\phi_{ab}}/(2\sqrt{\kappa_a\kappa_b})$, where the sign of the phase comes from the convention in Eq. \ref{eq:H_2modes}. The corresponding graph representation is shown in Fig.~\ref{figSItheory}(b). When comparing to the EoMs for frequency conversion \eqref{eq:eom_fc}, one notices three differences: (1) the mode amplitude $a$ is now coupled to the conjugate mode amplitude $b^{*}$, (2) the complex detuning term for the conjugate mode is conjugated and with a minus sign, and (3) the coupling term for the conjugate mode has a minus sign. These subtle differences lead to the following scattering matrix:
\begin{equation}
\mathbf{S}=\left(\begin{array}{ll}
\dfrac{i\eta_a\Delta_b^*}{\Delta_a\Delta_b^*-\abs{\beta_{ab}}^2}-1 & \dfrac{i\sqrt{\eta_a\eta_b}\beta_{ab}}{\Delta_a\Delta_b^*-\abs{\beta_{ab}}^2} \\
\dfrac{-i\sqrt{\eta_a\eta_b}\beta_{ab}^*}{\Delta_a\Delta_b^*-\abs{\beta_{ab}}^2} & \dfrac{-i\eta_b\Delta_a}{\Delta_a\Delta_b^*-\abs{\beta_{ab}}^2}-1
\end{array}\right).
\label{eq:S_PA}
\end{equation}
On resonance ($\Delta_j=i/2$) and neglecting internal loss ($\eta_j=1$) we recover Eq.~\eqref{eq:S_PA_0}:
\begin{equation*}
\mathbf{S}=\left(\begin{array}{cc}
\dfrac{1+4\abs{\beta_{ab}}^2}{1-4\abs{\beta_{ab}}^2} & \dfrac{4i\beta_{ab}}{1-4\abs{\beta_{ab}}^2} \\
\dfrac{-4i\beta_{ab}^*}{1-4\abs{\beta_{ab}}^2} & \dfrac{1+4\abs{\beta_{ab}}^2}{1-4\abs{\beta_{ab}}^2}
\end{array}\right).
\end{equation*}
The scattering parameters diverge for $\abs{\beta_{ab}}=1/2$, in stark contrast with the frequency conversion case. In the limit $\abs{\beta_{ab}}\rightarrow1/2^-$ each scattering parameter has an amplitude gain, $\sqrt{G}\approx2/(1-4\abs{\beta_{ab}}^2)$. Similar to the frequency conversion case, we observe $S_{ba} \neq S_{ba}$ for $\beta_{ab}$ complex.

\subsubsection{Three-mode systems}
Parametric frequency conversion and parametric amplification are the building blocks from which we can construct complex coupled-mode networks. Both processes exhibit a directional phase shift, which is the first ingredient to build nonreciprocal devices like isolators and circulators. The second ingredient required is an interferometer to unambiguously define the reference phase. Here we will build such an interferometer in frequency space by parametrically coupling three modes to form a loop. We will use the matrix formalism and its graph counterpart, which become very useful when extending the mode basis to three or more modes. Indeed, the coupling network is well captured by the mode-coupling matrix $\mathbf{M}$ alone. Let us now consider three oscillators with resonant frequencies $\omega_{j}$ and loss rates $\kappa_j$, driven at the signal frequencies $\omega_{j}^\text{s}$, for $j\in \{a,b,c\}$.

\paragraph{Circulation}
To build a circulator we will connect these three modes via frequency conversion. For that, we simultaneously modulate the coupling between the modes at the difference frequencies $\omega_{ab}^p\approx\abs{\omega_b-\omega_a}$, $\omega_{bc}^p\approx\abs{\omega_c-\omega_b}$, and $\omega^{p}_{ac}\approx\abs{\omega_a-\omega_c}$, satisfying the condition $\omega_{ab}^p + \omega_{bc}^p = \omega_{ac}^p$. The resulting equations of motion for the internal modes $\mathbf{A}=(a,b,c)^T$ are described by the mode-coupling matrix
\begin{equation}
\mathbf{M}=
\begin{pmatrix}
\Delta_{a} & \beta_{ab} & \beta_{ac}\\
\beta_{ab}^{*} & \Delta_{b} & \beta_{bc} \\
\beta_{ac}^* & \beta_{bc}^{*} & \Delta_{c}
\end{pmatrix},
\end{equation}
where $\Delta_{j}=(\omega_{j}^\text{s}-\omega_{j})/\kappa_{j}+i/2$ is the complex detuning for mode $j$, and  $\beta_{jk}=|g_{jk}|e^{i\phi_{jk}}/(2\sqrt{\kappa_j\kappa_k})$ is the normalized coupling between the modes $j$ and $k$. The graph representation of this Langevin matrix is shown in Fig.~\ref{figSItheory}(e), which highlights the loop structure of the coupling network. Each pair of modes is now coupled via two different paths. This leads to interferences, controlled by the loop phase $\phi_\mathrm{loop}=\phi_{ab}+\phi_{bc}-\phi_{ac}$. Solving for the scattering matrix using Eq.~\eqref{eq:eom_S} we obtain
\begin{widetext}
	\begin{equation}
	\mathbf{S}=\left(\begin{array}{lll}
	i\eta_a\dfrac{\Delta_b\Delta_c-\abs{\beta_{bc}}^2}{\abs{\mathbf{M}}}-1 &
	i\sqrt{\eta_a\eta_b}\dfrac{\beta_{ac}\beta_{bc}^*-\beta_{ab}\Delta_c}{\abs{\mathbf{M}}} & 
	i\sqrt{\eta_a\eta_c}\dfrac{\beta_{ab}\beta_{bc}-\beta_{ac}\Delta_b}{\abs{\mathbf{M}}} \\
	i\sqrt{\eta_a\eta_b}\dfrac{\beta_{ac}^*\beta_{bc}-\beta_{ab}^*\Delta_c}{\abs{\mathbf{M}}} &
	i\eta_b\dfrac{\Delta_a\Delta_c-\abs{\beta_\text{ac}}^2}{\abs{\mathbf{M}}}-1 & 
	i\sqrt{\eta_b\eta_c}\dfrac{\beta_{ab}^*\beta_{ac}-\beta_{bc}\Delta_a}{\abs{\mathbf{M}}} \\
	i\sqrt{\eta_a\eta_c}\dfrac{\beta_{ab}^*\beta_{bc}^*-\beta_{ac}^*\Delta_b}{\abs{\mathbf{M}}} &
	i\sqrt{\eta_b\eta_c}\dfrac{\beta_{ab}\beta_{ac}^*-\beta_{bc}^*\Delta_a}{\abs{\mathbf{M}}} & 
	i\eta_c\dfrac{\Delta_a\Delta_b-\abs{\beta_{ab}}^2}{\abs{\mathbf{M}}}-1
	\end{array}\right),
	\label{eq:S_Circ}
	\end{equation}
\end{widetext}
where $\abs{\mathbf{M}}=\Delta_a\Delta_b\Delta_c-\abs{\beta_{bc}}^2\Delta_a - \abs{\beta_{ac}}^2\Delta_b - \abs{\beta_{ab}}^2\Delta_c + 2\abs{\beta_{ab}}\abs{\beta_{bc}}\abs{\beta_{ac}}\cos\phi_\mathrm{loop}$. Importantly, compared to the two-mode cases, the \textit{magnitudes} of the scattering parameters are now nonreciprocal,  $\abs{S_{ij}}\neq \abs{S_{ji}}$. Maximum isolation from mode $b$ to mode $a$ ($S_{ab}=0$) is achieved for $\beta_{ac}\beta_{bc}^*=\beta_{ab}\Delta_c$. On resonance ($\Delta_{j}=i/2$) and neglecting internal loss ($\eta_{j}=1$), this condition coincides with unity transmission from $a$ to $b$, $\abs{S_{ba}}=1$ for $\abs{\beta_{ab}}=\abs{\beta_{bc}}=\abs{\beta_{ac}}=1/2$ and for a loop phase $\phi_\mathrm{loop}=-\pi/2$. Setting $\phi_{ab}=\phi_{bc}=-\phi_{ac} = \pi/2$, one recovers the ideal circulator scattering matrix from Eq.~\eqref{eq:S_Circulator_0}:
\begin{equation*}
\mathbf{S}=\left(\begin{array}{ccc}
0 & 0 & 1 \\
1 & 0 & 0 \\
0 & 1 & 0 
\end{array}\right).
\end{equation*}

\begin{figure*}
	\includegraphics[scale=1]{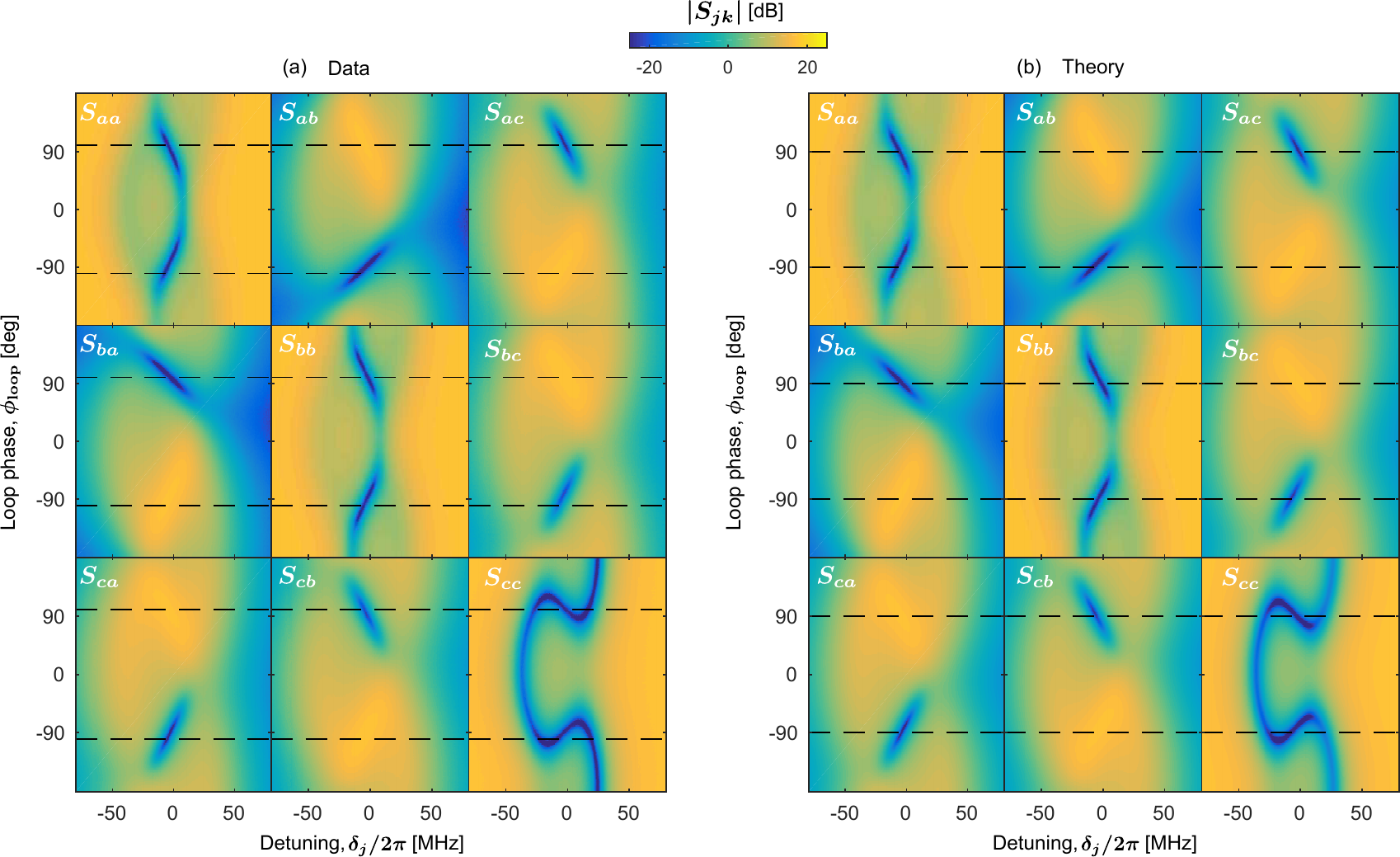}
	\caption{Circulator. Magnitude of the measured (a), and simulated (b), scattering parameters as a function of the loop phase $\phi_\mathrm{loop}$ for near-ideal pump strengths.
		\label{figSIcirculator}}
\end{figure*}

In Fig.~\ref{figSIcirculator} we show the measured and simulated scattering parameters as a function of the total loop phase $\phi_\mathrm{loop}$ for the FPJA operating as a circulator. Excellent agreement is found between data and theory. For the opposite loop phase $\phi_\mathrm{loop}=\pi/2$, the direction of circulation is reversed. 

\paragraph{Directional phase-preserving amplification}

\begin{figure*}
	\includegraphics[scale=1]{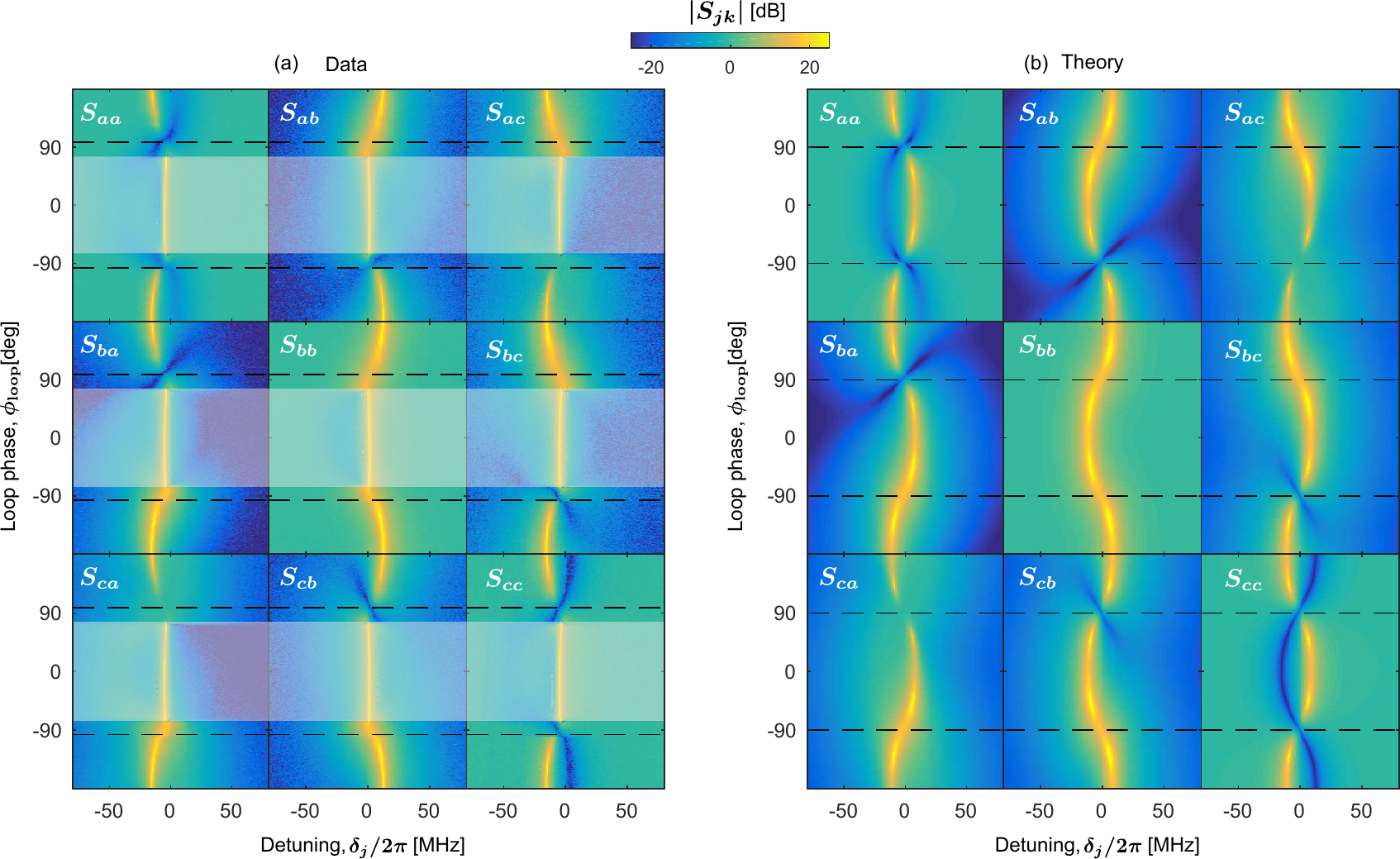}
	\caption{Directional phase-preserving amplifier. Magnitude of the measured (a), and simulated (b), scattering parameters as a function of the loop phase $\phi_\mathrm{loop}$ for near-ideal pump strengths. In practice the device undergoes free oscillations for loop phase $\abs{\phi_\mathrm{loop}}\lesssim80$ degrees (see Appendix B.3).
		\label{figSIdelta}}
\end{figure*}

To build a directional amplifier we will connect two pairs of modes via amplification and close the interferometer using frequency conversion. For that, we simultaneously modulate the coupling between the modes at the frequencies $\omega_{ab}^p\approx\omega_a^\text{s}+\omega_b^\text{s}$, $\omega_{bc}^p\approx\omega_b^\text{s}+\omega_c^\text{s}$, and $\omega_{ac}^p\approx\abs{\omega_a^\text{s}-\omega_c^\text{s}}$, satisfying the condition $\omega_{ac}^p + \omega_{ab}^p = \omega_{bc}^p$. The resulting equations of motion for the internal modes $\mathbf{A}=(a,b^*,c)^T$ are described by the matrix
\begin{equation}
\mathbf{M}=
\begin{pmatrix}
\Delta_{a} & \beta_{ab} & \beta_{ac}\\
-\beta_{ab}^{*} & -\Delta^*_{b} & \beta_{bc} \\
\beta_{ac}^* & -\beta_{bc}^{*} & \Delta_{c}
\end{pmatrix}, \label{eq:DeltaAmpM}
\end{equation}
The detuning and coupling terms for the frequency conversion and amplification processes are defined using the same convention than in the two-mode cases, with $\Delta_{j}=(\omega_{j}^\text{s}-\omega_{j})/\kappa_{j}+i/2$, $\beta_{ac}=|g_{ac}|e^{i\phi_{ac}}/(2\sqrt{\kappa_a\kappa_c})$ and $\beta_{jk}=|g_{jk}|e^{-i\phi_{jk}}/(2\sqrt{\kappa_j\kappa_k})$, where ${j},{k} \in \{{b,c}\}$. The graph representation of this mode-coupling matrix is shown in Fig.~\ref{figSItheory}(f), forming again an interferometer. Solving for the scattering matrix using Eq.~\eqref{eq:eom_S} we obtain
\begin{widetext}
	\begin{equation}
	\mathbf{S}=\left(\begin{array}{lll}
	i\eta_a\dfrac{\abs{\beta_{bc}}^2-\Delta_b^*\Delta_c}{\abs{\mathbf{M}}}-1 &
	i\sqrt{\eta_a\eta_b}\dfrac{-\beta_{ab}\Delta_c-\beta_{ac}\beta_{bc}^*}{\abs{\mathbf{M}}} & 
	i\sqrt{\eta_a\eta_c}\dfrac{\beta_{ac}\Delta_b^*+\beta_{ab}\beta_{bc}}{\abs{\mathbf{M}}}  \\
	i\sqrt{\eta_a\eta_b}\dfrac{\beta_{ab}^*\Delta_c+\beta_{ac}^*\beta_{bc}}{\abs{\mathbf{M}}}  &
	i\eta_b\dfrac{\Delta_a\Delta_c-\abs{\beta_{ac}}^2}{\abs{\mathbf{M}}}-1 & 
	i\sqrt{\eta_b\eta_c}\dfrac{-\beta_{bc}\Delta_a-\beta_{ab}^*\beta_{ac}}{\abs{\mathbf{M}}} \\
	i\sqrt{\eta_a\eta_c}\dfrac{\beta_{ac}^*\Delta_b^*+\beta_{ab}^*\beta_{bc}^*}{\abs{\mathbf{M}}} &
	i\sqrt{\eta_b\eta_c}\dfrac{\beta_{bc}^*\Delta_a+\beta_{ab}\beta_{ac}^*}{\abs{\mathbf{M}}} & 
	i\eta_c\dfrac{\abs{\beta_{ab}}^2-\Delta_a\Delta_b^*}{\abs{\mathbf{M}}}-1
	\end{array}\right),
	\label{eq:S_Delta}
	\end{equation}
\end{widetext}
where $\abs{\mathbf{M}}=-\Delta_a\Delta_b^*\Delta_c + \abs{\beta_{bc}}^2\Delta_a + \abs{\beta_{ac}}^2\Delta_b^* + \abs{\beta_{ab}}^2\Delta_c + 2\abs{\beta_{ab}}\abs{\beta_{bc}}\abs{\beta_{ac}} \cos\phi_\mathrm{loop} $ and $\phi_\mathrm{loop}=\phi_{ab}+\phi_{bc}+\phi_{ac}$. Similar to the circulation case, the \textit{magnitude} of the scattering parameters are nonreciprocal. The condition for maximum isolation from port $b$ to port $a$, $S_{ab}=0$, is achieved for $\beta_{ac}\beta_{bc}^*=-\beta_{ab}\Delta_c$. On resonance, this condition leads to a loop phase $\phi_\mathrm{loop}=-\pi/2$. We further neglect internal loss and require an input match $\abs{S_{aa}}=0$ on resonance, which can be accomplished by choosing the matching condition for the frequency conversion branch ($\abs{\beta_{ac}}=1/2$) and letting the two amplification branches be equal $\abs{\beta_{ab}}=\abs{\beta_{bc}}$. To fix the phases of the scattering matrix elements, we choose $\phi_{ab}=\phi_{ac}=-\phi_{bc}=-\pi/2$ to arrive at the scattering matrix of Eq.~\eqref{eq:S_Delta_0}:
\begin{equation*}
\mathbf{S}=\left(\begin{array}{ccc}
0 & 0 & 1 \\
\sqrt{G-1} & \sqrt{G} & 0 \\
\sqrt{G} & \sqrt{G-1}& 0 
\end{array}\right),
\end{equation*}
where $\sqrt{G} = (1+4\abs{\beta_{ab}}^2)/(1-4\abs{\beta_{ab}}^2)$. Let's consider a coherent signal  at the input of mode $a$, while modes $b$ and $c$ are seeded by vacuum fluctuations. Both the signal in mode $a$ and the vacuum fluctuations in mode $b$ are amplified towards modes $b$ and $c$. This leads to the same minimum system noise as a standard two-mode amplifier. Finally, the vacuum fluctuations in $c$ are routed to mode $a$ with unity gain. The measured and simulated scattering parameters as a function of the total loop phase $\phi_\mathrm{loop}$ for FPJA operation as a directional amplifier are shown in Fig.~\ref{figSIdelta}. Excellent agreement is found between data and theory for $\abs{\phi_\mathrm{loop}}\gtrsim80$. For $\abs{\phi_\mathrm{loop}}\lesssim80$ degrees the system undergoes free oscillations. While for a two-mode amplifier (see Eq. \ref{eq:S_PA}) the free oscillation threshold is simply $\abs{\beta_{ab}}=1/2$, the threshold for a three-mode directional amplifier becomes phase dependent. A general method to calculate if the system will free oscillate is to look for signal frequencies solutions of $\abs{\mathbf{M}}=0$ and with a positive imaginary part.  

\subsubsection{Generalization to multiple input/output ports per modes}
In this section we generalize the formalism to include multiple input ports for each mode, crucial for accurate noise calculation in next section.

In section B.1 to B.3 we have considered each mode $j$ to be coupled to a single external port and accounted for internal loss using the coupling efficiency parameter $\eta_{j}=\kappa^\text{ext}_{j}/\kappa_{j}$. This is sufficient for calculating the scattering parameters between the external ports of interest. However, to describe system with multiple external (or internal) ports per mode, and to reach accurate noise calculations, one need to generalize this formalism. Intuitively, if a fraction of the input signal can be lost in the environment, then, conversely, noise from this environment enters the system. In other word, in order to preserve the commutators of the input and output fields, one must account for all the ports contributing to the total loss rates of each mode.

The generalization to multiple ports per mode consist in a straightforward redefinition of the input and ouput field, $\mathbf{A}_\text{in}$ and $\mathbf{A}_\text{out}$, and of the matrix $\mathbf{H}$. Note that because the network of coupled internal modes stays unchanged, so does the mode-coupling matrix $\mathbf{M}$. Each mode $j$ can be coupled to multiple ports (which can include ``ports'' to the thermal environment), so that $\kappa_{j} = \sum_k\kappa_j^k$, where $k$ indexes all ports to which mode $j$ is coupled. For a system of $n$ modes coupled to a total of $m$ ports,  $\mathbf{A}_\text{in}$ and $\mathbf{A}_\text{out}$ become vectors of length $m$, $\mathbf{H}$ becomes a matrix of size $n\times m$ where $H_{jk}=\sqrt{\eta_j^k}$, and $\mathbf{M}$ remains a matrix of size $n\times n$. The condition $\sum_k\eta_j^k=1$ ensures that all noise sources are accounted for. Eq.~\eqref{eq:eom_S} becomes 
\begin{equation}
\mathbf{S} = i\mathbf{H}^T\mathbf{M}^{-1}\mathbf{H}-\mathds{1}. \label{eq:S_gen}
\end{equation}

As an example, we consider the directional amplifier shown in B.3.a and include a single additional port per mode to describe the coupling to the environment at a rate $\kappa_j^\text{int}$, so that $\kappa_j=\kappa_j^\text{ext}+\kappa_j^\text{int}$. We define the input field as $\mathbf{A_\text{in}}=(\hat{a}_\text{in},\hat{\xi}_{a,\text{in}},\hat{b}^\dagger_\text{in},\hat{\xi}^\dagger_{b,\text{in}},\hat{c}_\text{in}, \hat{\xi}_{c,\text{in}})^T$ where $\hat{\xi}_{j,\text{in}}$ is the input field for the environment of mode $j$. Similarly we define $\mathbf{A_\text{out}}=(\hat{a}_\text{out},\hat{\xi}_{a,\text{out}},\hat{b}^\dagger_\text{out},\hat{\xi}^\dagger_{b,\text{out}},\hat{c}_\text{out}, \hat{\xi}_{c,\text{out}})^T$. The matrix $\mathbf{H}$ becomes:
\begin{equation*}
\mathbf{H}=\left(\begin{array}{cccccc}
\sqrt{\eta_a^\text{ext}} & \sqrt{\eta_a^\text{int}} & 0 & 0 & 0 & 0 \\
0 & 0 & \sqrt{\eta_b^\text{ext}} & \sqrt{\eta_b^\text{int}} & 0 & 0 \\
0 & 0 & 0 & 0 & \sqrt{\eta_c^\text{ext}} & \sqrt{\eta_c^\text{int}} \\
\end{array}\right).
\end{equation*}

\subsubsection{Output noise}
In this section we use the generalized scattering parameters (Eq. \ref{eq:S_gen}) to calculate the output noise of a system of parametrically coupled mode. The quantum noise spectral density of the output fields \cite{Clerk2010}, $\mathcal{N}[\omega]$, is defined as
\begin{equation}
2\pi\mathcal{N}[\omega]\delta(\omega-\omega') =\left<\mathbf{A}_\text{out}^\dagger[\omega]\mathbf{A}_\text{out}^T[\omega']\right>, \label{eq:PSDfromcov}
\end{equation}

where $\left<\mathbf{A}_\text{out}^\dagger[\omega]\mathbf{A}_\text{out}^T[\omega']\right>/2\pi \delta(\omega-\omega')$ is the covariance matrix of the output field, in units of photons. One can then express the spectral density of the output field using Eq. \ref{eq:PSDfromcov} and the relation between input and output covariance matrices:
\begin{equation}
\left<\mathbf{A}_\text{out}^\dagger[\omega']\mathbf{A}_\text{out}^T[\omega]\right> = \mathbf{S}^*[\omega']\left<\mathbf{A}_\text{in}^\dagger[\omega']\mathbf{A}_\text{in}^T[\omega]\right> \mathbf{S}^T[\omega]. \label{eq:S_cov}
\end{equation}

The input covariance matrix contains all information about the input noise, like bath temperatures or correlations, and can describe both classical and non-classical noise source. For example, if the environment of mode $j$ is in a thermal state, the corresponding input noise follows $\left<\hat{\xi}_{j,\text{in}}^\dagger[\omega']\hat{\xi}_{j,\text{in}}[\omega]\right> = 2\pi n_{j,\text{th}}\delta(\omega-\omega')$ and $\left<\hat{\xi}_{j,\text{in}}[\omega]\hat{\xi}_{j,\text{in}}^\dagger[\omega']\right> = 2\pi (n_{j,\text{in}}+1)\delta(\omega-\omega')$, where $n_{j,\text{th}} = (\exp(\hbar\omega_j/k_BT) - 1)^{-1}$ is the mean thermal occupation number at the resonator frequency $\omega_j$ at a temperature $T$ ($k_B$ is the Boltzmann constant).

Note that in this work, we perform linear measurements of the output field quadratures. Therefore we access the symmetrized (classical) spectral density \cite{Clerk2010}, $\bar{\mathcal{N}}[\omega]$:  
\begin{equation}
\bar{\mathcal{N}}[\omega] = \frac{1}{2}\left(\mathcal{N}[\omega]+\mathcal{N}[-\omega]\right)
\end{equation}

The theoretical predictions for the output noise made in Figs \ref{figPA}(e), \ref{figDelta}(f) and \ref{figDelta}(g) are based on our best guess for the scattering parameters, shown in \ref{figPA}(d) and \ref{figDelta}(d), and assume the each port is seeded by vacuum noise.

\subsubsection{Discussion}
The agreement between theory and experiment is primarily due to a clean mode basis. Indeed, the EoMs rely on the rotating wave approximation to eliminate dynamics at any spurious frequencies. If the approximation cannot be made, new modes appear in the coupled network, leading for example to losses or added noise. The device presented here was carefully designed so that all first order parametric processes would be separated in frequency. In Fig.~\ref{figSItheory}(a) we show the mode frequencies $\omega_{j}$ and all the modulation frequencies $\abs{\omega_{j}\pm\omega_{k}}$ for ${j},{k} \in \{{a,b,c}\}$ as a function of flux. The shaded areas represent a bandwidth three times larger than the largest mode width ($3\kappa_\text{c}/2\pi=180~\text{MHz}$), necessary to ensure a good rotating wave approximation. At the flux bias chosen in this work, $\Phi/\Phi_0\approx0.29$, all these processes are well separated.

\section*{Appendix C: System noise calibration}

\begin{figure*}
	\includegraphics[scale=1]{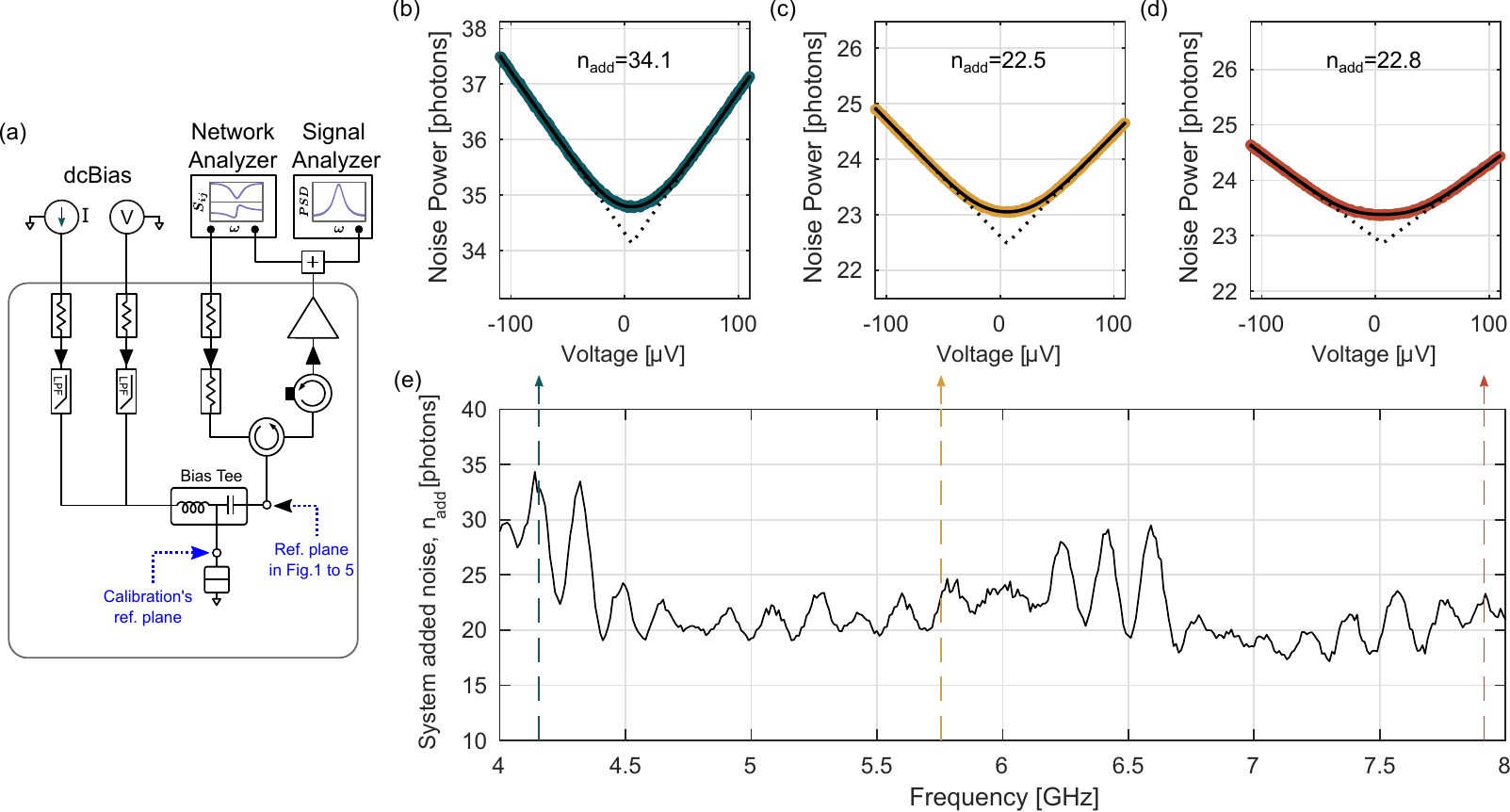}
	\caption{Noise Calibration. (a) The noise emitted by a biased metallic tunnel junction is measured using the same setup as for the FPJA measurement in Fig. \ref*{figSetup}(a). (b,c,d) Measured power spectral density in photon unit as a function of the voltage across the shot-noise junction, measured at the frequencies $\omega_{1}/2\pi=4.155~\mathrm{GHz}$, $\omega_{2}/2\pi=5.756~\mathrm{GHz}$ and $\omega_{3}/2\pi=7.915~\mathrm{GHz}$, corresponding to the resonances of the FPJA shown in Fig. \ref{figSetup}(c). The solid black lines are a fit to Eq. \ref{eq:shotnoise}. The dashed are linear extrapolation of the noise at high voltage, and their crossing point correspond the added noise (e) Measured system added noise as a function of frequency.
		\label{figSIShotNoise}}
\end{figure*}

The calibration of the noise performance of an amplifier at milli-Kelvin temperature and close to the standard quantum limit is not a trivial task. Most commonly used is the so-called Y-factor method. It requires a calibrated noise source, such as a variable-temperature resistor \cite{Kindel2016}, a circuit QED system \cite{Macklin2015}, or a biased metallic tunnel junction \cite{Roch2012,Spietz2003}.  While each technique is subject to various experimental and conceptual challenges, they all share a common problem: they calibrate the system added noise to a reference plane that is usually not the one of the amplifier. For a simplistic but concrete and general example, consider that the noise source and the amplifier are connected by a transmission line with unknown loss: the calibrated noise at the output of the source is now uncalibrated at the input of the amplifier. This leads to the following general consideration when calibrating the noise of an amplifier: all components necessary for the proper operation of the amplifier (circulators, filters, couplers, cables, etc.) should be included in the system noise calibration.

In this work we choose a slightly different approach. In a separate cooldown we replace the FPJA by a metallic tunnel junction and a bias tee, leaving every other component of the measurement chain identical (including the HEMT amplifier bias parameters), as shown in Fig. \ref{figSIShotNoise}(a). This allows for the calibration of the system noise down to the junction's reference plane, which only differs from the FPJA reference plane by the loss of the bias tee ($<0.5~\mathrm{dB}$). We therefore obtain an upper bound for the system noise temperature at the reference plane of the FPJA.

The power spectral density of the noise emitted by a metalic tunnel junction, at a frequency $\omega$ and temperature $T$, as a function of the voltage $V$ across the junction is $\mathcal{N}=\mathcal{N}_++\mathcal{N}_-$ (unit of $\mathrm{quanta\cdot s^{-1}\cdot Hz^{-1}}$), where

\begin{equation}
\mathcal{N}_\pm=\frac{k_BT}{2\hbar\omega}\left[\frac{eV\pm\hbar\omega}{2k_BT}\coth\left(\frac{eV\pm\hbar\omega}{2k_BT}\right)\right],
\end{equation}

$k_B$ is the Boltzmann constant, $e$ is the electron charge and $\hbar$ is the reduced Planck constant. The gain of the full measurement chain, $G_\text{sys}$, and the noise \textit{added} by that chain, $n_\text{add}$ (unit of $\mathrm{quanta}$), is extracted from the measured power spectral density, $\mathcal{N}_\text{meas}$, following:

\begin{equation}
\mathcal{N}_\text{meas}= G_\text{sys}\left(\mathcal{N}+n_\text{add}\right) \label{eq:shotnoise}
\end{equation}

The power spectral density in photon units, $\mathcal{N}_\text{meas}/G_\text{sys}$,  measured at the frequencies $\omega_{1}/2\pi=4.155~\mathrm{GHz}$, $\omega_{2}/2\pi=5.756~\mathrm{GHz}$ and $\omega_{3}/2\pi=7.915~\mathrm{GHz}$, corresponding to the resonance of the FPJA, are shown respectively in Figs. \ref{figSIShotNoise}(b), \ref{figSIShotNoise}(c) and \ref{figSIShotNoise}(d). From a fit to Eq. \ref{eq:shotnoise}, we extract a constant temperature of $100~\mathrm{mK}$ and a system added noise of $n_\text{add}=34.1\mathrm{,}~22.5~\mathrm{and}~22.8$ respectively at $\omega_{1}$, $\omega_{2}$ and $\omega_{3}$. Similar measurements are performed over the $4~\mathrm{GHz}$ to $8~\mathrm{GHz}$ band and the measured system noise as a function of frequency is shown in Fig. \ref{figSIShotNoise}. One can see oscillations, due to slight impedance mismatches and component imperfections throughout the chain. The temperature is consistently measured at $100~\mathrm{mK}$, probably due to imperfect thermalisation of the sample box.

\bibliographystyle{plain}

\end{document}